\documentclass[12pt]{article}
\usepackage{amsmath}
\usepackage{amssymb}
\usepackage{hyperref}
\usepackage{graphicx,psfrag,epsf}
\usepackage{enumerate}
\usepackage{natbib}
\usepackage{url} 
\usepackage{latexsym,amsfonts}
\usepackage{float}
\newcommand{\blind}{0}
\usepackage[usenames,dvipsnames]{color}
\usepackage{caption}
\usepackage{subcaption}
\newtheorem{assumption}{Assumption}

\usepackage{chngcntr}

\def\frac#1#2{{\textstyle{#1\over#2}}}

\DeclareSymbolFont{AMSb}{U}{msb}{m}{n}
\DeclareMathSymbol{\Natural}{\mathbin}{AMSb}{"4E}
\DeclareMathSymbol{\Integer}{\mathbin}{AMSb}{"5A}
\DeclareMathSymbol{\Real}{\mathbin}{AMSb}{"52}
\DeclareMathSymbol{\Rational}{\mathbin}{AMSb}{"51}
\DeclareMathSymbol{\Imaginary}{\mathbin}{AMSb}{"49}
\DeclareMathSymbol{\Complex}{\mathbin}{AMSb}{"43} 
\DeclareMathSymbol{\Disk}{\mathbin}{AMSb}{"44} 
\def\bi{\begin{itemize}}
\def\ei{\end{itemize}}
\def\bd{\begin{description}}
\def\ed{\end{description}}
\def\ben{\begin{enumerate}}
\def\een{\end{enumerate}}
\def\bv{\begin{verbatim}}
\def\ev{\end{verbatim}}

\def\hat#1{{\widehat{#1}}}

\newcommand{\bs}{\boldsymbol}

\def\E{{\rm E}}

\def\2to{{\ {\buildrel 2\over \longrightarrow}\ }}

\def\iid{{\ {\buildrel \rm{iid}\over \sim}\ }}
\def\ind{{\ {\buildrel \rm{ind}\over \sim}\ }}

\def\I1ton{{$I_1,\ldots,I_n$}}
\def\X1ton{{$X_1,\ldots,X_n$}}
\def\Y1ton{{$Y_1,\ldots,Y_n$}}
\def\Z1ton{{$Z_1,\ldots,Z_n$}}
\def\R1ton{{$R_1,\ldots,R_n$}}
\def\e1ton{{$e_1,\ldots,e_n$}}
\def\t1ton{{$t_1,\ldots,t_n$}}
\def\x1ton{{$x_1,\ldots,x_n$}}
\def\y1ton{{$y_1,\ldots,y_n$}}
\def\z1ton{{$z_1,\ldots,z_n$}}

\begin{document}

\def\spacingset#1{\renewcommand{\baselinestretch}%
{#1}\small\normalsize} \spacingset{1}

\if1\blind
{
  \title{\bf Title}
  \author{Author 1\thanks{
    The authors gratefully acknowledge \textit{please remember to list all relevant funding sources in the unblinded version}}\hspace{.2cm}\\
    Department of YYY, University of XXX\\
    and \\
    Author 2 \\
    Department of ZZZ, University of WWW}
  \maketitle
} \fi

\if0\blind
{
  \bigskip
  \bigskip
  \bigskip
  \begin{center}
    {\LARGE\bf Causal Spatial Quantile Regression}
\end{center}

\begin{center}
\large 
Yan Gong$^{1}$, Reetam Majumder$^2$, Brian J. Reich$^3$, Rapha\"el Huser$^4$
\end{center}
\baselineskip=12pt
\vskip 5mm

\footnotetext[1]{
\baselineskip=10pt Harvard T.H. Chan School of Public Health, Boston, Massachusetts, United States. \\E-mail: yangong@hsph.harvard.edu}
\footnotetext[2]{
\baselineskip=10pt  Department of Mathematical Sciences, University of Arkansas, Fayetteville, Arkansas, United States. E-mail: reetamm@uark.edu}
\footnotetext[3]{
\baselineskip=10pt Department of Statistics, North Carolina State University, Raleigh, North Carolina, United States. E-mail: bjreich@ncsu.edu}
\footnotetext[4]{
\baselineskip=10pt Statistics Program, CEMSE Division, King Abdullah University of Science and Technology, Thuwal, Saudi Arabia. E-mail: raphael.huser@kaust.edu.sa}

\baselineskip=17pt
\vskip 4mm
\centerline{\today}
\vskip 6mm
\begin{center}
{\large{\bf Abstract}}
\end{center}
Treatment effects in a wide range of economic, environmental, and epidemiological applications often vary across space, and understanding the heterogeneity of causal effects across space and outcome quantiles is a critical challenge in spatial causal inference. To effectively capture spatial heterogeneity in distributional treatment effects, we propose a novel semiparametric neural network-based causal framework leveraging deep spatial quantile regression and then construct a plug-in estimator for spatial quantile treatment effects (SQTE). This framework incorporates an efficient adjustment procedure to mitigate the impact of spatial hidden confounders. Extensive simulations across various scenarios demonstrate that our methodology can accurately estimate SQTE, even with the presence of spatial hidden confounders. Additionally, the spatial confounding adjustment procedure effectively reduces neighborhood spatial patterns in the residuals. We apply this method to assess the spatially varying quantile treatment effects of maternal smoking on newborn birth weight in North Carolina, United States. Our findings consistently show negative effects across all birth weight quantiles, with particularly severe impacts observed in the lower quantile regions.

} \fi



\baselineskip=16pt

\par\vfill\noindent

\pagenumbering{arabic}
\baselineskip=26pt
\newpage
\spacingset{1.45} 

\newpage
\section{Introduction}\label{sec:introduction}
Identifying heterogeneity in treatment effects is crucial across various fields, including economic, environmental, and epidemiological applications. These applications span topics such as optimal customer targeting policies \citep{hitsch2024heterogeneous, yang2024targeting}, personalized medicine \citep{kent2018personalized, venkatasubramaniam2023comparison}, air pollution epidemiology \citep{de2024ambient, zorzetto2024confounder}, and climate-induced disease risks at the intersection of animal, human, and environmental domains \citep{rocklov2023decision}. Machine learning methods have proven effective in estimating heterogeneous treatment effects. For instance, \cite{athey2015machine} and \cite{wager2018estimation} demonstrated that their non-parametric causal forest algorithm is pointwise consistent for the true treatment effects. However, within their framework, it is assumed that treatment effects exhibit heterogeneity solely in the mean of the response population.


To achieve a more comprehensive understanding of treatment effects, it is beneficial to employ methods that estimate these effects across the entire distribution of the response population. In this context, the quantile treatment effect (QTE) emerges as a valuable estimation target. In economic applications, instrumental variable quantile regression (IVQR) models, introduced by \cite{chernozhukov2005iv, chernozhukov2006instrumental} and further developed by \cite{wuthrich2019closed}, serve as flexible plug-in estimators. These models utilize closed-form solutions derived from IVQR moment conditions to target QTE in the presence of endogenous covariates. Building on the potential outcomes framework \citep{rubin1974estimating}, \cite{sun2021causal} propose a general estimator for the population-level weighted QTE and provide a comprehensive review of relevant environmental applications. Additionally, \cite{firpo2007efficient} offers an efficient semiparametric method for estimating QTE, while \cite{xu2018bayesian} introduces a Bayesian nonparametric approach for causal inference on quantiles in the presence of confounders. More recently, \cite{xu2022bayesian} advanced this field by introducing a Bayesian semiparametric method for estimating QTE.

In the context of spatial causal inference \citep{reich2021review, gao2022causal, akbari2023spatial}, identifying causal effects is particularly challenging due to the presence of spatial confounding \citep{gilbert2021causal}, which necessitates additional assumptions. Further complicating statistical modeling and inference are the spatial correlations between treatment and response variables. Additionally, issues such as spillover or interference \citep{giffin2023generalized, papadogeorgou2023spatial} violate standard assumptions crucial for identifying causal effects, presenting another layer of complexity.

In recent advancements in spatial quantile regression, notable developments include the application of spatial panel quantile models to analyze the quantile co-movement structure of the U.S. stock market \citep{ando2023spatial}, spatial quantile autoregression models to examine the spatiotemporal evolution of daily maximum temperatures in Spain \citep{castillo2023spatial}, spatial geo-additive quantile regression models to explore the determinants of Airbnb prices in New York City \citep{bernardi2023determinants}, and quantile spatially varying coefficient model (QSVCM) which accounts for spatial nonstationarity over complex or irregular domains \citep{kim2025estimation}. While these spatial regression models demonstrate significant applicability, they typically focus on individual quantiles, requiring additional constraints to ensure non-crossing properties. For the joint modeling of the whole spatial quantile curve, foundational works have been done for spatial quantile regression \citep{reich2011bayesian, chen2021joint} and spatio-temporal quantile regression \citep{reich2012spatiotemporal}.

In this paper, we introduce a pioneering causal framework for identifying spatial heterogeneity in distributional treatment effects. Our approach features a flexible semiparametric neural network-based deep spatial quantile regression model, designed to capture complex dependencies among high-dimensional variables and intricate covariate interactions. We develop a plug-in estimator for spatial quantile treatment effects (SQTE) that can evaluate treatment effects that are non-linear, spatially varying, and quantile-dependent. Additionally, we propose a spatial confounding adjustment procedure to further reduce biases in both outcome regression and effect estimation. 

The remainder of this paper is organized as follows: Section~\ref{sec:methodology} introduces our novel causal spatial quantile regression approach. We start by introducing the deep spatial quantile regression model. Following this, we detail the assumptions necessary for identifying SQTE, describe the construction of a plug-in estimator, and present our method for spatial confounding adjustment. Section~\ref{sec:simulation} reports the findings from an extensive simulation study, exploring scenarios involving 1) no confounders; 2) observed (spatial and non-spatial) confounders; and 3) hidden (spatial and non-spatial) confounders. We evaluate the model's performance in predicting responses and estimating SQTE. In Section~\ref{sec:app}, we apply our methodology to quantify the spatially varying treatment effects of maternal smoking on the birth weight distribution of newborns in North Carolina, USA, with a focus on the impacts on newborns with low birth weights.

\section{Methodology}\label{sec:methodology}

 Throughout, we assume that we have observational data $\mathcal{O} = (Y, T, \bs X^T, \bs s^T)^T$, where $Y\in \mathbb{R}$ is the response variable of interest, $T\in\mathcal{T}$ is the treatment variable,  $\bs X = (X_1,\ldots, X_d)^T\in \mathbb{R}^d$ is a $d$-dimensional random vector of covariates, and $\bs s\in \bs S \subset \mathcal{R}^2$ denotes the spatial coordinates of the observations. Note that the observed covariates $\bs X$ and treatment variable $T$ can be spatially-dependent or spatially-independent.

\subsection{Spatial quantile modeling}
Understanding the heterogeneity of causal effects across space and outcome quantiles is a critical challenge in spatial causal inference. To address this, we introduce a framework for spatial quantile modeling that extends the potential outcomes paradigm to quantify treatment effects that vary spatially and depend on specific quantiles of the response distribution. Our approach integrates spatial dependencies and distributional heterogeneity by merging two advanced methodologies: DeepKriging \citep{chen2020deepkriging}, a deep neural network architecture optimized for spatial mean prediction, and Semiparametric Quantile Regression (SPQR) \citep{xu2021bayesian, xu2022spqr}, a flexible framework that models conditional quantiles using monotonic splines with neural network-weighted components.

By synthesizing DeepKriging’s capacity to capture complex spatial structures with SPQR’s ability to estimate quantile-specific relationships, we develop a unified model that accounts for spatial autocorrelation and reveals how causal effects differ across quantiles. This integration provides a powerful tool for uncovering nuanced, location-specific variations in treatment impacts, particularly in settings where effects may disproportionately influence different segments of the outcome distribution. Within this approach, the conditional density and cumulative distribution functions of the response variable $Y$ can be approximated by


$$
f(y\mid T,\bs X,\bs s) = \sum_{k=1}^K \theta_k(T,\bs X,\bs s) M_k(y),
$$
and
$$
F(y\mid T,\bs X,\bs s) = \sum_{k=1}^K \theta_k(T,\bs X,\bs s) I_k(y),
$$
respectively, where the $M_k(y)$ terms are $K$ second-order $M$-spline basis functions, the $I_k(y)$ terms are $K$ second-order $I$-spline basis functions with equally-spaced knots on the span $[0,1]$, and $\bs s= (s_1, s_2)^T$ are the spatial coordinates. Note that $I$-splines are defined as the integral of $M$-splines, which have to be positive and integrate to one (to be a valid density function) in addition to other necessary conditions for splines. Given the conditional cumulative distribution function, the conditional quantile function at quantile level $\tau\in[0,1]$ can be obtained as the inverse
\begin{equation}\label{equ:q}
    q(\tau \mid T,\bs X,\bs s) = F^{-1}(\tau \mid T,\bs X,\bs s).
\end{equation}

Moreover, the mixture weights $\theta_k(T,\bs X,\bs s)$ must satisfy the following conditions:
$$
\theta_k(T,\bs X,\bs s) \geq0, \quad \sum_{k=1}^K \theta_k(T,\bs X,\bs s) = 1.
$$
To ensure flexibility and satisfy these constraints, the weights are specified as an $L$-layer feed-forward neural network with a softmax output activation function, i.e.,
$$
\theta_k(T,\bs X,\bs s) = \dfrac{\exp\{z_k^{(L)}(T,\bs X,\bs s, \mathcal{W})\}}{\sum_{k=1}^K\exp\{z_k^{(L)}(T,\bs X,\bs s, \mathcal{W})\}},
$$
where $\mathcal{W}=\{\bs W^{(1)},\ldots, \bs W^{(L)}\}$ a set of neural network weight matrices, with $\bs W^{(l)}\in \mathbb{R}^{(V_l+1)\times V_{l-1}}$ and $V_l$ the number of neural units, while $z_k^{(l)}$ denotes the output functions at layer $l= 1,\ldots,L$. Neural network weights $\mathcal{W}$ often arise from a multi-layer perceptron architecture. 

To further incorporate structured spatial information into our model and capture spatially varying quantile effects, we construct spatial covariates that represent important spatial features at multiple resolutions \citep{nychka2015multiresolution}. Given spatial locations $\bs s$, we can generate multi-resolution radial basis functions at $M$ resolutions,
$$\bs \phi(\bs s) = \{\bs\phi_1(\bs s),\ldots,\bs\phi_M(\bs s)\},\quad m=1,\ldots,M,$$ 
where the set of basis functions at the $m-$th resolution has the form $$\bs\phi_m(\bs s) = \{\varphi(||\bs s -\bs u_j||/\delta_m); j=1,\ldots,p\}^T$$ for some bandwidth parameter $\delta_m>0$, such that $\delta_1>\delta_2>\cdots>\delta_M$, and $\{\bs u_j; j = 1,\ldots, p\}$ are $p$ node points placed 
on a rectangular grid covering the spatial domain of interest. Here, we use compactly supported basis functions from the Wendland family, which may be expressed as:
\begin{equation}\label{eq:bf}
\varphi(h) = (1-h^6)(35h^2+18h+3)/3, \quad h\in[0,1], 
\end{equation}
and $\varphi(h) = 0,$ otherwise. The number of basis functions $p$ at each resolution level should typically depend on the choice of bandwidth $\delta_m$, and guidelines can be found in \cite{chen2020deepkriging}. Then we obtain the extended expressions of the conditional density and cumulative distribution functions of $Y$ with the mixture weights as $\theta_k(T,\bs X,\bs s, \bs \phi(\bs s))$, for $k = 1,\ldots,K$.




\subsection{Identification of spatial quantile treatment effects}
Under the potential outcome framework \citep{neyman1923applications,rubin1974estimating}, we further assume that the treatment assignment is binary, i.e., $\mathcal{T} = \{0,1\}$, for simplicity, where $T=1$ indicates active treatment. The response variable $Y$ of each ``unit" is only obtained under a single treatment $T\in\mathcal{T}$, so the response that would have been obtained under the (alternative) treatment $T\in\{0,1\}$ is usually called a ``potential outcome" and written $Y(T), T = 0,1.$ Estimating the treatment effect is difficult due to the fact that one observes $Y(0)$ only or $Y(1)$ only, but this is made possible under certain (often untestable) underlying assumptions. These include the following:

\begin{assumption}[Stable Unit Treatment Value Assumption, SUTVA]
1) There is one version of the treatment. 2) No interference: units do not interfere with each other; the potential outcomes for one unit are independent of the treatment assignment of others.
\end{assumption}
\begin{assumption}[Consistency] For any unit, the observed outcome given assigned treatment is equal to the potential outcome.
\end{assumption}
\begin{assumption}[Ignorability] The treatment assignment is independent of the potential outcomes given the confounders.
(also referred to as a ``no hidden confounders" assumption.)
\end{assumption}
\begin{assumption}[Positivity] For any given value of confounders, the probability of assignment to active treatment is strictly between $(0, 1)$.
\end{assumption}


\noindent In the context of this paper, we shall assume that all these assumptions hold. While Assumptions 1,2, and 4 are often quite realistic, Assumption 3 is usually violated and untestable in real data applications. We will test the model's performance when there are hidden confounders in our simulation study. Given Assumptions 1--4, the spatially varying $\tau$-th quantile treatment effects can be defined as 
$$
\Delta(\tau,\bs s) = \mathbb{E}_{\bs X}\big\{q(\tau \mid T :=1,\bs X,\bs s)- q(\tau \mid T :=0,\bs X,\bs s)\big\},
$$
where $q(\tau|\cdot)$ is the quantile function in Equation~(\ref{equ:q}), and we write for simplicity as 
$$
\Delta(\tau,\bs s) = \mathbb{E}_{\bs X}\big\{q^1(\tau\mid \bs X,\bs s)- q^0(\tau\mid \bs X,\bs s)\big\} = \int_\mathcal{\bs X}\big\{q^1(\tau\mid \bs X,\bs s)- q^0(\tau\mid \bs X,\bs s)\big\}\nu(d \bs X),
$$
for some probability measure $\nu(\cdot)$. Assume we have $\mathcal{O}_i = (Y_i, T_i, \bs X_i, \bs s_p)^T, i=1,\ldots, n_p$, observations on $P$ unique locations, where the number of observations $n_p$ at each location can be different, for $p = 1,\ldots,P$. A plug-in estimator can be built to estimate $\Delta(\tau,\bs s_p)$:
$$
\hat{\Delta}(\tau,\bs s_p) = \dfrac{1}{n_p}\sum_{i=1}^{n_p}\big\{\hat{q^1}(\tau\mid \bs X_i,\bs s_p)- \hat{q^0}(\tau\mid \bs X_i,\bs s_p)\big\},
$$
for $\tau\in(0,1)$. The spatially-averaged QTE at quantile $\tau$ can be expressed as
$$
\Delta(\tau) = \mathbb{E}_{\bs s}\big[\mathbb{E}_{\bs X}\big\{q^1(\tau\mid \bs X,\bs s)- q^0(\tau\mid \bs X,\bs s)\big\}\big] = \int_\mathcal{\bs s}\int_\mathcal{\bs X}\big\{q^1(\tau\mid \bs X,\bs s)- q^0(\tau\mid \bs X,\bs s)\big\}\nu(d \bs X)\mu(d \bs s),
$$
for some probability measure $\nu(\cdot)$ and $\mu(\cdot)$. In the data application presented in Section~\ref{sec:app}, we have a particular interest in the lower tail of the birth weight, and the corresponding estimator can be obtained as
$$
\hat{\Delta}(\tau) = \dfrac{1}{P}\sum_{p=1}^P\big\{\hat{\Delta}(\tau,\bs s_p)\}.
$$



\subsection{Neighborhood spatial confounding adjustment}\label{sec:neighborhoodadjustment}
The proposed modeling approach is adept at capturing intricate non-spatial and spatial relationships among covariates, treatment variables, and the response. When confounding variables are observable, the model inherently accounts for them by incorporating all covariates in its formulation, thus negating the need for additional confounding adjustments. Conversely, unobserved confounders can impact the outcome and introduce bias in estimating causal effects unless proper adjustments are made. 

When hidden confounders exhibit spatial correlation, adjustment methods from recent literature can be implemented, including matching \citep{jarner2002estimation}, spatial propensity score methods \citep{davis2019addressing}, and bias mitigation as discussed by \cite{schnell2020mitigating}. For a comprehensive review of these methods, we can refer to \cite{reich2021review}. In this work, we introduce a spatial confounding adjustment procedure to address the presence of potential hidden confounders around the neighborhood, which is particularly effective when the hidden confounders are smoother than the treatment variable in space. To estimate the SQTE at a specific location $\bs s_p, p = 1,\ldots, P$, we selectively use a subset of observations positioned within a predetermined distance from $\bs s_p$. This selection ensures that a certain proportion of the total observations are incorporated into the regression analysis, thereby maintaining statistical robustness. The distance is determined empirically, and the reference point for the initial subregional model fitting is chosen to be approximately the center of the spatial region. Then we use the estimated coefficients to construct the predictions under different treatment scenarios for the whole spatial domain. By estimating effects locally, we can remove confounding bias caused by variables that vary smoothly over space and are thus roughly constant for all observations in each local fit. Furthermore, we include a comprehensive sensitivity analysis of the application study to assess the model's performance under varying distance criteria. Detailed results of this analysis are provided in the Appendix.



\section{Simulation study}\label{sec:simulation}
To demonstrate our method's performance in predicting the response and estimating heterogeneous treatment effects, we set up a simulation study that includes both non-spatial and spatial covariates, as well as random effects. We generate data in three scenarios, where there are either 1) no confounders, 2) observed confounders, or 3) hidden confounders. The confounders are either spatial or non-spatial.

\subsection{General setup}
Spatial data are simulated on a regular grid over the unit square $[0,1]^2$, where the number of locations is set to $n_s = 20^2 = 400$. At each location, we generate $n=1000$ observations from five different models and three scenarios, and we then use our methodology to predict responses and estimate the SQTE. The simulations are conducted $N$ =100 times for each model and scenario to quantify the uncertainty.

We simulate six types of observed (i.e., non-hidden) covariates, which are either ``spatial" or ``non-spatial" (i.e., spatially indexed but independent and identically distributed (i.i.d.) across the space), and we denote them in short-hand notation as $\bs X_k(\bs s_l) = (X_{1k}(\bs s_l), X_{2k}(\bs s_l),$
$\ldots, X_{6k}(\bs s_l))^T$, for $k=1,\dots n, l=1,\ldots,n_{s}$. Specifically, covariates are simulated as follows:
\begin{align*} 
 X_{1k}(\bs s_l) &\iid \text{Unif}(0, 1),&\quad k=1,\dots,n,l=1,\ldots,n_s,\\
 X_{2k}(\bs s_l) &\iid \mathcal{N}(0, 1),&\quad k=1,\dots,n,l=1,\ldots,n_s,\\
 X_{3k}(\bs s_l) &\iid \text{Binomial}(1, 0.5),&\quad k=1,\dots,n,l=1,\ldots,n_s,\\
 \{X_{4k}(\bs s_1),X_{4k}(\bs s_2),\ldots,X_{4k}(\bs s_{n_s})\}^T &\iid\mathcal{N}_{n_s}(\bs 0, \Sigma(0.1)),&\quad k=1,\dots,n,\\
\{X_{5k}(\bs s_1),X_{5k}(\bs s_2),\ldots,X_{5k}(\bs s_{n_s})\}^T &\iid\mathcal{N}_{n_s}(\bs 0, \Sigma(0.2)),&\quad k=1,\dots,n,\\
\{X_{6k}(\bs s_1),X_{6k}(\bs s_2),\ldots,X_{6k}(\bs s_{n_s})\}^T &\iid\mathcal{N}_{n_s}(\bs 0, \Sigma(0.5)),&\quad k=1,\dots,n,
\end{align*}
where $\mathcal{N}_{n_s}(\bs 0, \Sigma(\lambda))$ denotes the multivariate normal distribution stemming from a stationary spatial Gaussian process with zero mean and squared-exponential covariance function $\exp\{-(h/\lambda)^2\}$ observed at $n_s$ grid locations, where $h$ is the distance between locations and $\lambda>0$ is the range parameter.
Note that the first three covariates are non-spatial (i.e., spatially independent), 
while the last three are spatial (i.e., spatially dependent) and are distributed according to independent Gaussian processes. Moreover, we simulate three types of hidden non-spatial and spatial variables which are denoted as $\bs H_k(\bs s_l) = \{H_{1k}(\bs s_l), H_{2k}(\bs s_l), H_{3k}(\bs s_l)\}^T$, for $k=1,\dots n, l=1,\ldots,n_{s}$, where
\begin{align*} 
 H_{1k}(\bs s_l) &\iid \mathcal{N}(0, 1),&\quad k=1,\dots,n,l=1,\ldots,n_s,\\
\{H_{2k}(\bs s_1),H_{2k}(\bs s_2),\ldots,H_{2k}(\bs s_{n_s})\}^T &\iid\mathcal{N}_{n_s}(\bs 0, \Sigma(0.4)),&\quad k=1,\dots,n,\\
 H_{3k}(\bs s_l) &= \sin(5\pi s_{l1}) + \cos(2\pi s_{l2}),&\quad k=1,\dots,n,l=1,\ldots,n_s.
\end{align*}

Furthermore, the binary treatment variable is assumed to be spatially varying and is simulated as $T(\boldsymbol{s}_l)\iid{\rm Ber}(p(\boldsymbol{s}_l))$, where $p(\boldsymbol{s}_l)\in(0,1)$ is the probability of receiving the treatment at location $\boldsymbol{s}_l, l=1,\ldots,n_s,$ and may depend on observed or hidden confounders. We assume that all $n_s$ variables at location $\bs s_l$ receive the same treatment. We also consider a scenario where the treatment is non-spatial, see details in Section~\ref{sec:unconfoundedness}. 
Then the data-generating mechanism is
$$
q_{\boldsymbol{Y}_k}(\bs s_l) = f\{\bs X_k(\bs s_l)\} + g\{\bs s_l, T(\bs s_l)\} + h\{\bs H_k(\bs s_l)\} +q_{\epsilon_k}(\bs s_l),
$$
with some functions $f\{\cdot\}$, $g\{\cdot\}$ and $h\{\cdot\}$ to be specified, $q(\cdot)$ is the quantile function, and random residual effects are assumed to have the form $\epsilon_k(\bs s_l) \ind \mathcal{N}(\text{sin}(2\pi s_{l1})+\text{cos}(3\pi s_{l2}),1)$, for $k=1,\dots,n, l=1,\ldots,n_{s}$.
\subsection{Models considered}\label{sec:5models}
In the simulation study, we consider five different models that include all the observed covariates and the treatment variable within each model. When there are hidden confounders, we denote the models with confounding adjustments in the model name by AD. The models may be summarized as follows:
\begin{itemize}
    \item Model 1: Covariates only;
    \item Model 2: Covariates + Spatial coordinates;
    \item Model 3: Covariates + Spatial features (one resolution, $M = 1$);
    \item Model 4: Covariates + Spatial features (two resolutions, $M = 2$);
    \item Model 5: Covariates + Spatial features (three resolutions, $M = 3$).
\end{itemize}
The spatial features are defined by the Wendland spatial basis functions in (\ref{eq:bf}). At each resolution level $m=1,\ldots, M$, we specify the number of basis functions to be $p=3^2, 5^2$, and $7^2$ in the two-dimensional space. We follow the same principles as in \cite{chen2020deepkriging} for the choice of scale parameter $\delta_m$ and implementation details.
The data are split into training sets (80\%) and testing sets (20\%) and we evaluate the prediction and estimation performance of each model by computing the root mean integrated squared error (RMISE) of the predicted response and the estimated QTE on a logarithmic scale at different quantile levels, where the results are averaged over all spatial locations. We further present the spatial patterns of the mean (of 100 simulation results) RMISE at the quantile level $\tau = 0.05$ (with standard deviation shown in the Appendix for predicting the response and estimating SQTE over the spatial domain.

\subsection{Simulation scenarios and results}\label{sec:simulationscenarios}
We consider three scenarios, namely data generated under unconfoundedness, with observed confounders, and with hidden confounders. Here, we provide more details on each scenario and discuss the results of our simulation study.

\subsubsection{Scenario 1: Unconfoundedness}\label{sec:unconfoundedness}
Adopting a generic notation where, e.g., $\bs X(\bs s) = (X_{1}(\bs s), X_{2}(\bs s), \ldots, X_{6}(\bs s))^T$ (similarly for $T(\bs s), \bs H(\bs s)$, and $\epsilon(\bs s))$ denotes the location-specific covariate vector for a given replicate and $\bs s$ being an arbitrary location, we specify the form of the true quantile function when there are no hidden confounders as
\begin{align}
q_Y(\tau\mid T,\bs X,\bs s) =  & 3(\tau-1/2)\{X_1(\bs s)+3/5\}^3 + 5[X_2(\bs s) + 4\{X_2(\bs s)-1/2\}^2]\exp\{-X_2^2(\bs s)\}  \nonumber \\
&+2\exp[\{X_3(\bs s)+1/2\}^2\{X_4(\bs s)-1/2\}^2]\nonumber \\
& +6(\tau-1)\{X_5(\bs s)+2/5\} + 5\{X_6(\bs s)+1/2\}^2 \nonumber \\
&+ 2 s_1(\tau-1/2)^2 T(\bs s) \nonumber \\
&+ q_\epsilon(\bs s). \label{eq:true_quantile_noco} 
\end{align}
Therefore, the heterogeneous (non-linear, spatially varying, $\tau$-dependent) quantile treatment effect is $\Delta(\tau, \bs s) = 2 s_1(\tau-1/2)^2$. 
Under unconfoundedness, we assign the treatment probability as $p(\boldsymbol{s}) = 0.5$ (spatially constant) such that the treatment variable is not confounded by any observed or hidden variables. 
Figure~\ref{fig:p0_rmise} in the Appendix displays the log(RMISE) of predicted responses (left panel) and QTE estimates (right panel) at different quantile levels from the five models in Section~\ref{sec:5models}, illustrated by different colors, and plotted at different quantile levels. Since there are spatial random effects,  models with spatial features perform better in prediction than models that include only covariates. Even though some of the covariates are spatially varying, models without spatial features cannot accurately capture spatial random or fixed effects. The performance of our approach in estimating the QTE is, however, not significantly different among the models. We have better results for $\tau\approx0.5$, as expected. While it is harder to estimate the QTE for very low or very high $\tau$, the results indicate that the RMISE is nevertheless quite small across all quantile values considered here, even near the tails.
Furthermore, Figure~\ref{fig:p0_errmat} in the Appendix shows the spatial patterns of the mean RMISE values for predicting the response at the quantile level $\tau = 0.05$ from the five models. Spatial patterns exist for Models 1 and 2, and the RMISE is larger than for Models 3--5.
Finally, Figure~\ref{fig:p0_qfmat} in the Appendix shows the spatial patterns of RMISE for estimating SQTE over the spatial domain at the quantile level $\tau = 0.05$ from five models. Interestingly, spatial patterns exist for all models, and Model 3 has the lowest RMISE. This seems to indicate that including spatial features in the model is worthwhile. The true SQTE has a linear increasing trend in the direction of the longitude and we can also find longitudinal patterns in the RMISE from all the models.

\subsubsection{Scenario 2: With observed confounders}
In this scenario, we assign the treatment probability as $p(\bs s):= \text{expit}\{X_1(\bs s)+X_6(\bs s)\}$ such that the treatment is confounded by observed non-spatial and spatial covariates, where $\text{expit}(x) = {e^x}/({1 + e^x})$ is the inverse of the logit function, for $x\in\mathbb{R}$. The form of the true SQTE is the same as in (\ref{eq:true_quantile_noco}). 
Figure~\ref{fig:p1_rmise} in the Appendix displays the log(RMISE) of predicted responses (left panel) and QTE estimates (right panel) at different quantile levels from five models illustrated by different colors at different quantile levels. The prediction performances from the models are consistent with the results from Scenario 1 when there are no confounders. 
The spatial patterns in RMISE for predicting the response and estimating SQTE at the quantile level $\tau = 0.05$ are shown in Figure~\ref{fig:p1_errmat} and Figure~\ref{fig:p1_qfmat} in the Appendix, respectively. We find similar patterns, and the RMISE for both predicted responses and QTE estimates are on the same scale as in the unconfounding scenario. 
When there are confounding variables that are observed, including them explicitly in the regression model is an effective strategy, thanks to our flexible neural network-based regression framework, since the prediction error of the response and estimation error of the quantile treatment effects are comparable to Scenario 1. Furthermore, this also suggests that confounding adjustment is not necessary for this scenario.
\subsubsection{Scenario 3: With hidden confounders}
In this last scenario---the most complex and realistic one---we assume that there exist hidden confounders, and we specify the form of the true quantile function as
\begin{align*} 
q_Y(\tau\mid T,\bs X,\bs s) =  & 3(\tau-1/2)\{X_1(\bs s)+3/5\}^3 + 5[X_2(\bs s) + 4\{X_2(\bs s)-1/2\}^2]\exp\{-X_2^2(\bs s)\}  \\
&+2\exp[\{X_3(\bs s)+1/2\}^2\{X_4(\bs s)-1/2\}^2]\\
& +6(\tau-1)\{X_5(\bs s)+2/5\} + 5\{X_6(\bs s)+1/2\}^2 \\
&+3H_1^2(\bs s) + 3H_2(\bs s) + 5H_3(\bs s)\\
&+ 2s_1(\tau-1/2)^2 T(\bs s) \\
&+ q_\epsilon(\bs s),
\end{align*}
so that the heterogeneous (non-linear, spatially varying, $\tau$-dependent) quantile treatment effect remains the same as before. We assign the treatment probability as $p(\bs s):= \text{expit}\{5H_3(\bs s)\}$ such that the treatment is confounded by the hidden spatial variable $H_3(\bs s)$. This is the most challenging scenario that we consider in our simulation study.

\begin{figure}[t!]
\centering
\includegraphics[width=1\linewidth]{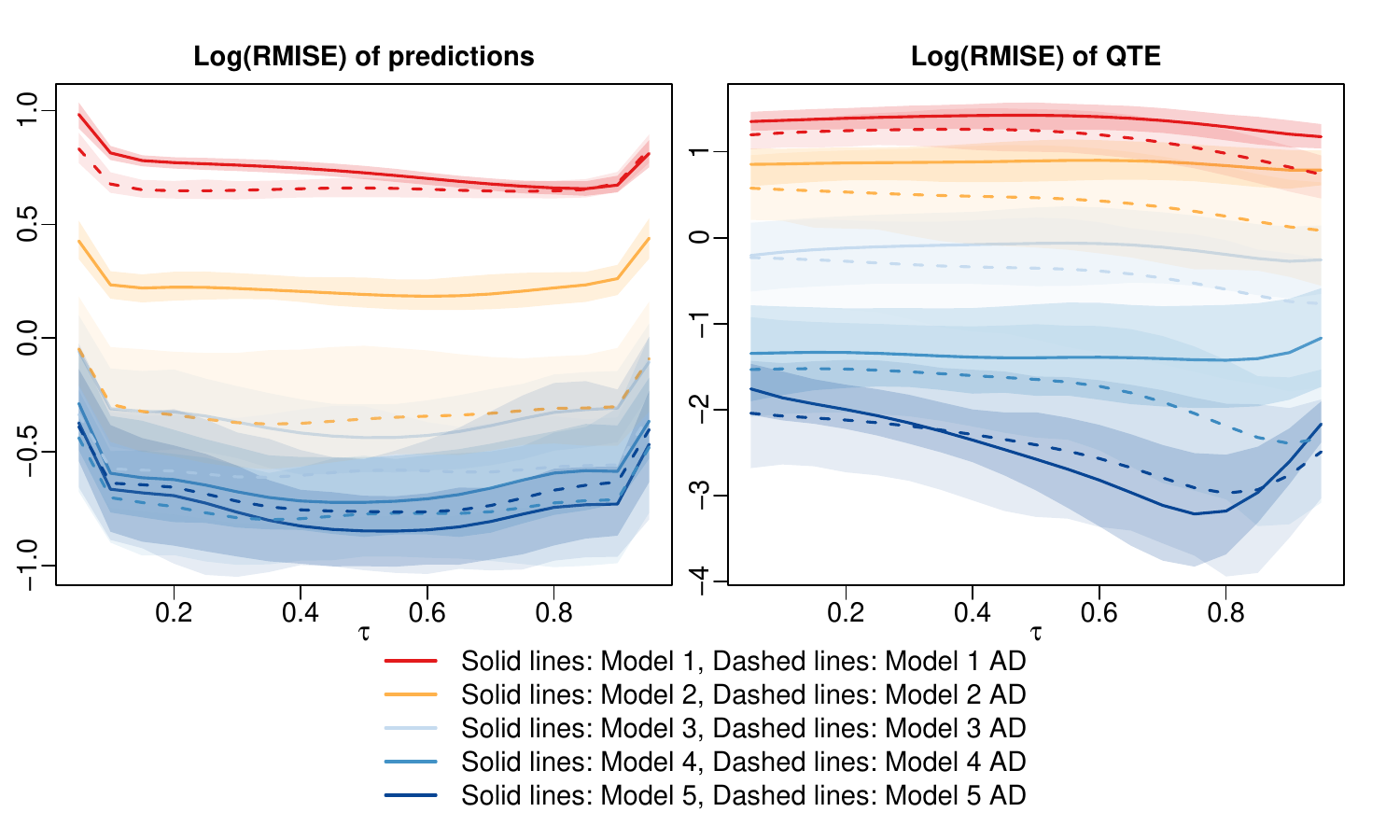}
\vspace{-1.1cm}
\caption{[Scenario 3: With hidden confounders.] Log(RMISE) of predicted responses (left panel) and QTE estimates (right panel) at different quantile levels from five models and the models with spatial confounding adjustment procedures (labeled as AD). The shaded areas correspond to 95\% confidence intervals from 100 simulations.}\label{fig:p3_rmise}
\end{figure}

Figure~\ref{fig:p3_rmise} shows the log(RMISE) of predicted responses (left panel) and QTE estimates (right panel) at different quantile levels from five models (displayed with different colors), plotted at the quantile level. The dashed lines display the results with the spatial confounding adjustment. In particular, we choose the empirical distance to include approximately 20\% of the whole dataset around the central location of the whole spatial domain. The basis functions are also created on the spatial locations that are within the subregion. Compared to the previous scenarios, the prediction errors are larger in all of the models since the data-generating process includes hidden confounders, which can be interpreted as random effects that are more important and potentially spatially structured. With regard to the estimation errors in QTE, we notice that the models' performances are generally quite different: specifically, Model 5 performs as well as in the unconfounding case, while the other models exhibit larger errors.

Furthermore, the spatial confounding adjustment appears to help reduce the RMISE in both predicted responses and QTE estimates for Models 1 and 2. However, when the spatial features are already accounted for in the model, the neighborhood spatial confounding adjustment does not reduce the bias effectively. Therefore, when highly structured confounders exist, the neighborhood spatial confounding adjustment may not be essential.

\begin{figure}[t!]
\centering
\includegraphics[width=1\linewidth]
{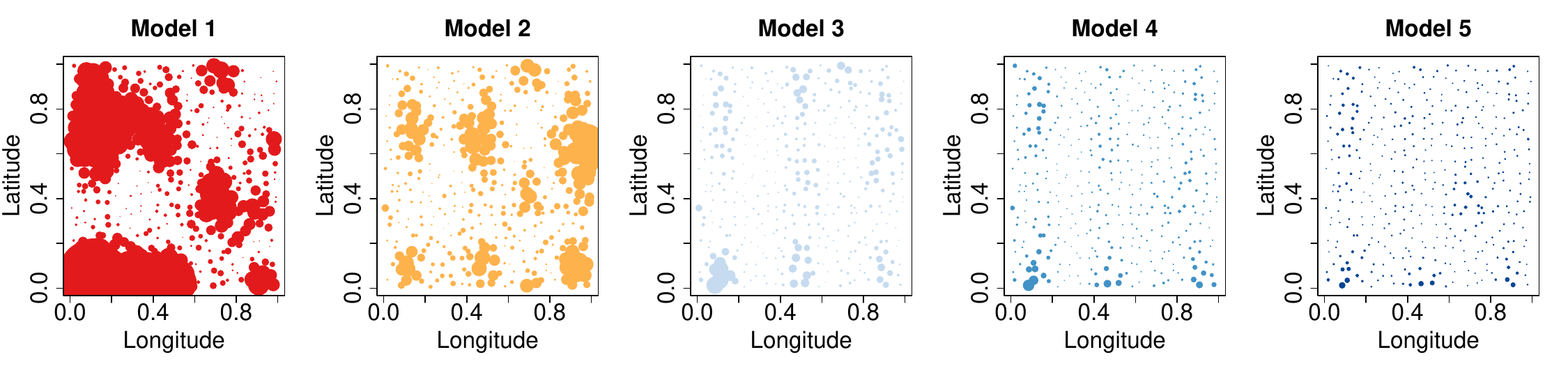}
\includegraphics[width=1\linewidth]{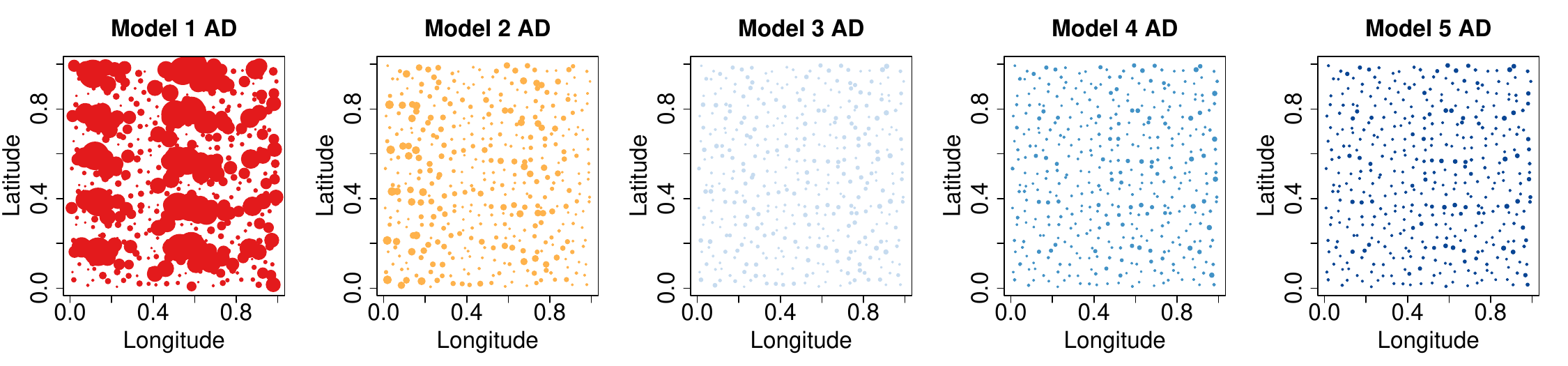}
\caption{[Scenario 3: With hidden confounders.] Spatial patterns of the mean (of 100 simulation results) RMISE for predicting the response at the quantile level $\tau = 0.05$ from the five models before (first row) and after (second row) spatial confounding adjustment. The size of the points is proportional to the RMISE values.}	\label{fig:p3_errmat}
\end{figure}

\begin{figure}[t!]
\centering
\includegraphics[width=1\linewidth]{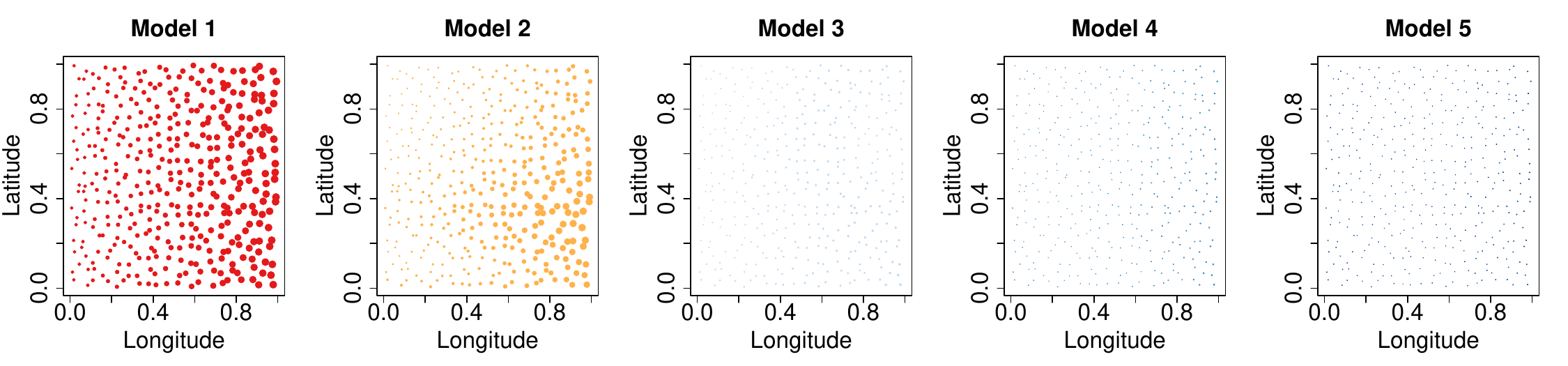}
\includegraphics[width=1\linewidth]{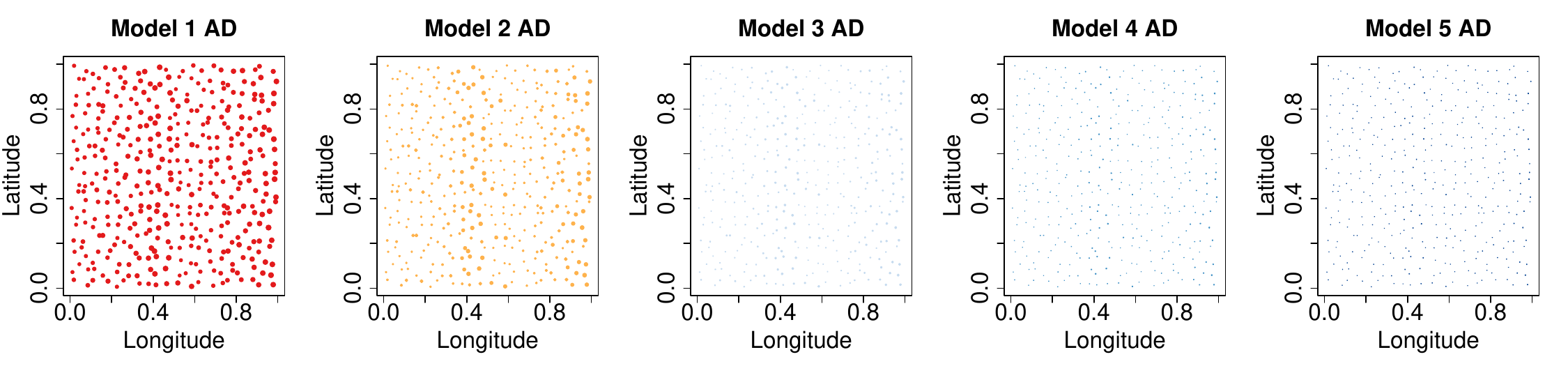}
\caption{[Scenario 3: With hidden confounders.] Spatial patterns of the mean (of 100 simulation results) RMISE for estimating SQTE at the quantile level $\tau = 0.05$ from the five models before (first row) and after (second row) spatial confounding adjustment. The size of the points is proportional to the RMISE values.}	\label{fig:p3_qfmat}
\end{figure}

Spatial patterns of the mean RMISE for predicting the response at the quantile level $\tau = 0.05$, computed based on 100 simulations, for the five models before and after spatial confounding adjustment are shown in Figure~\ref{fig:p3_errmat}. Due to the specific spatial confounding adjustment procedure that borrows neighborhood information in space, we can see that the RMISEs of all five models after the spatial adjustment are distributed more evenly, and the overall magnitude of the errors is decreased. 
Similar improvements in the RMISE of the SQTE estimates can be found in Figure~\ref{fig:p3_qfmat}.

\section{Data application}\label{sec:app}

\begin{figure}[t!]
	\centering
\includegraphics[width=1\linewidth]{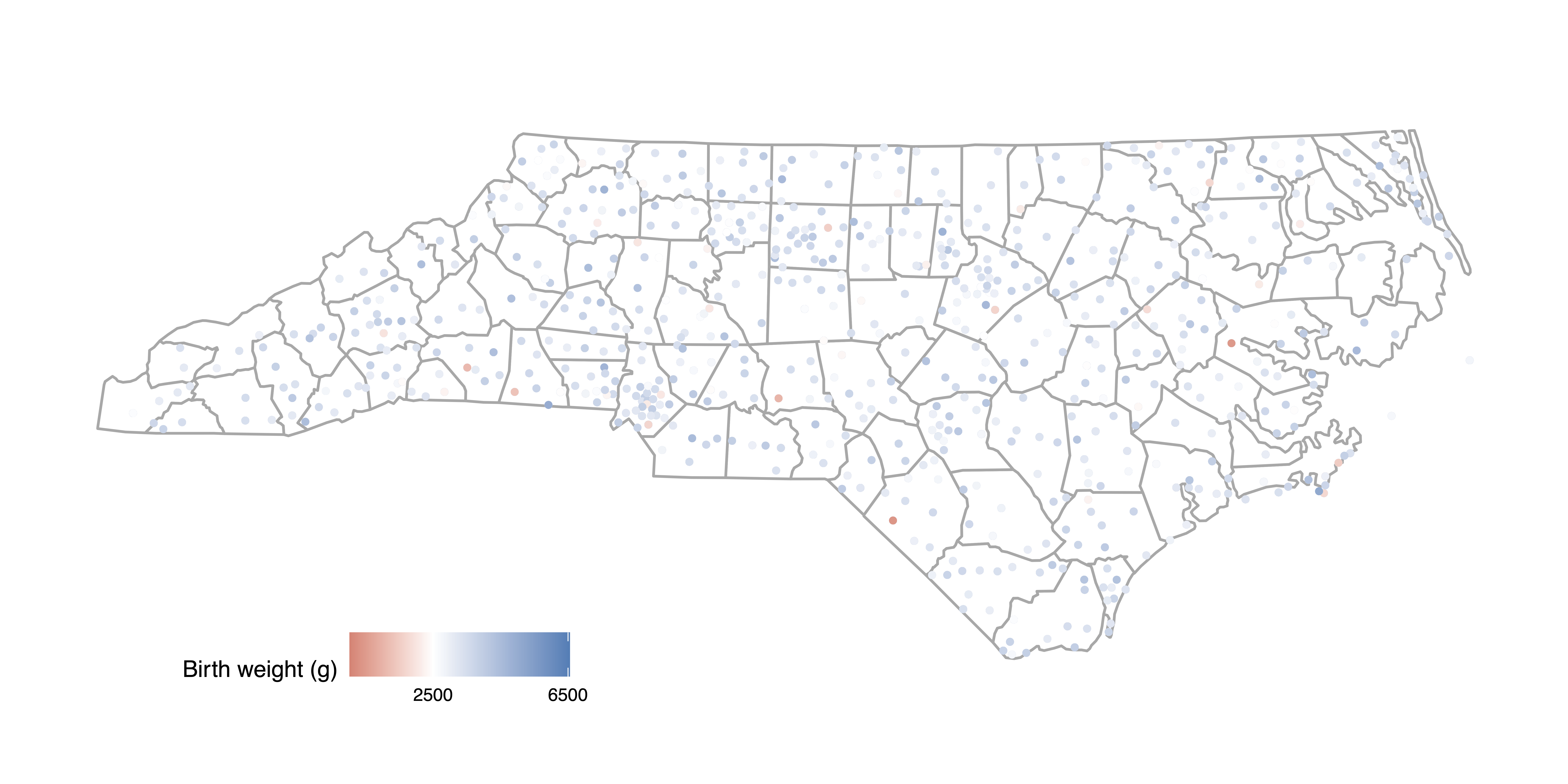}	
	\caption{Average birth weights (in grams) of newborns from first-time white mothers in North Carolina from 1988 to 2002. The color scale indicates the average birth weights of newborns at each location, and a birth weight less than 2,500 grams (5 lbs, 8 oz) is considered low birth weight (as shown in the red scale).}
	\label{fig:bw}
\end{figure}

We apply the proposed method to quantify spatial quantile treatment effects of maternal smoking on low birth weight of newborns in North Carolina, United States. We use a dataset that consists of records of newborns from 1988 to 2002 provided by the North Carolina State Center Health Services. This dataset has been analyzed previously by \cite{abrevaya2015estimating} to estimate conditional average treatment effects and by \cite{xu2022bayesian} to quantify (non-spatial) quantile treatment effects. The dataset consists of records of first-time mothers, including 157,989 black mothers and 433,558 white mothers. We use the latter subgroup of data with white mothers, as it has a larger number of records. The response variable $Y$ is the birth weight measured in grams of newborns, from first-time white mothers in North Carolina. The treatment variable $T$ is a binary variable indicating maternal smoking (1: Yes, 0: No). Our goal is to quantify spatially varying treatment effects of maternal smoking on the whole distribution of the birth weight of newborn babies, with particular attention to low birth weights. The geographical locations (ZIP code centroids) of the records of white mothers are shown in Figure~\ref{fig:bw}. The color scale indicates the average birth weights of newborns at each location, and a birth weight less than 2,500 grams (5 lbs, 8 oz) is considered low birth weight (as shown in the red scale). In total, there are 691 distinct spatial locations, and at each location, the number of observations ranges from 5 to 5593.


\begin{figure}[t!]
	\centering
    \includegraphics[width=1\linewidth]{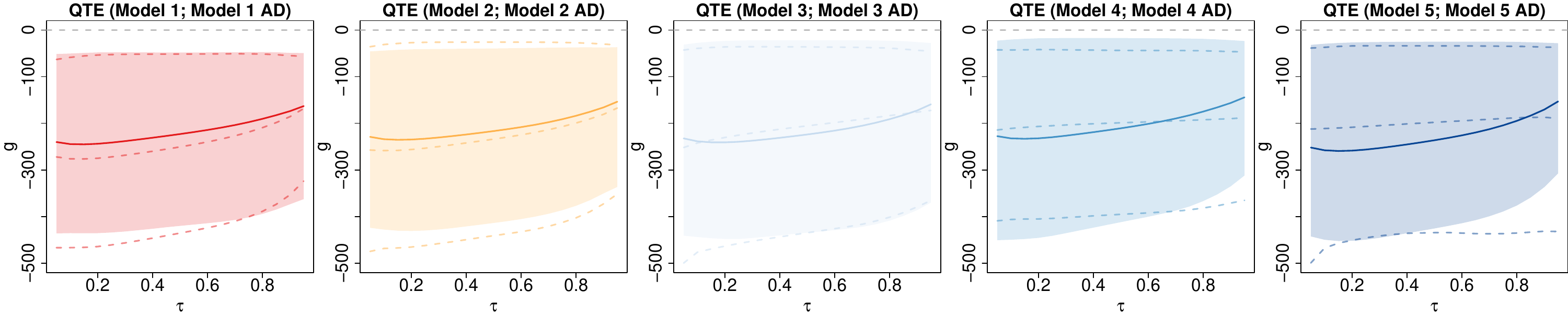}
	\caption{Estimated QTE (in grams) and 95\% confidence intervals from the five models before (solid lines and shaded areas) and after (dashed lines) spatial confounding adjustment at different quantile levels, averaged over all the spatial locations.}
	\label{fig:bw_qte}
\end{figure}

The five models (and their confounding-adjusted versions) that we used in the simulation study (recall Section~\ref{sec:5models}) are applied to estimate the quantile treatment effects and Figure~\ref{fig:bw_qte} shows the estimated QTE and 95\% confidence intervals from the five models before (solid lines and shaded areas) and after (dashed lines) spatial confounding adjustment at different quantile levels, averaged over all the spatial locations.
Consistently, the averaged quantile treatment effects are prone to be negative in all five models and exhibit slightly different trends of the effects at different quantile levels. A sensitivity analysis for the choice of coverage distance in spatial confounding adjustment is reported in the Appendix.

\begin{figure}[t!]
	\centering
\includegraphics[width=1\linewidth]{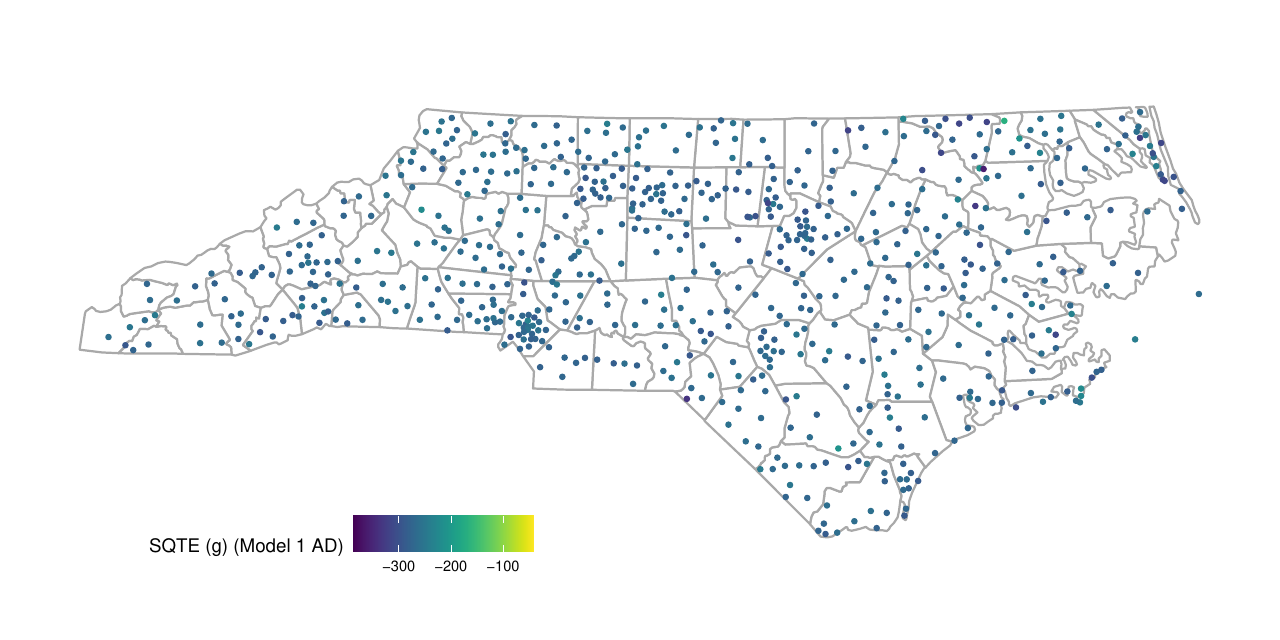}
 \includegraphics[width=1\linewidth]{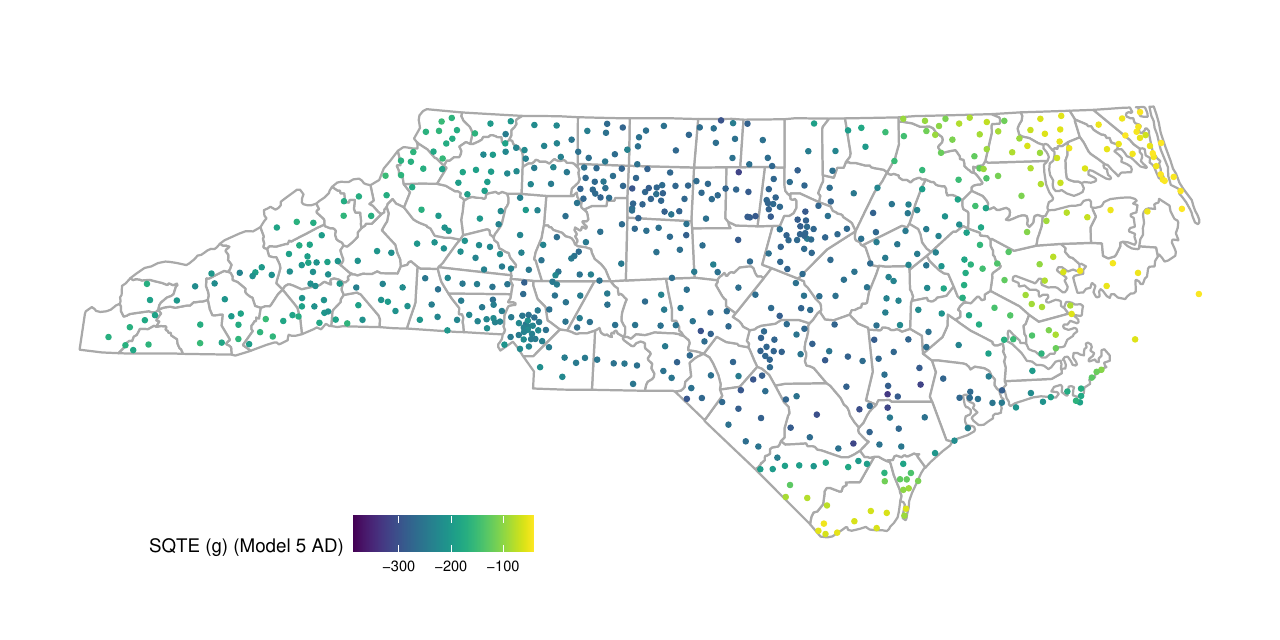}
	\caption{Estimated SQTE (in grams) of maternal smoking on the birth weight of newborn babies at $\tau = 0.05$ from Model 1 AD and Model 5 AD at all locations in North Carolina.}
	\label{fig:bw_sqte}
\end{figure}

To illustrate the spatial heterogeneity in the estimated quantile treatment effects, we further plot the estimated SQTE (in grams) at $\tau = 0.05$ from Model 1 AD and Model 5 AD in Figure~\ref{fig:bw_sqte}. By contrasting the two models, we find that the SQTE in the southern and northeastern parts of North Carolina is less homogeneous across space in Model 5 AD, and the negative effects of maternal smoking are slightly less significant but still exist. The estimated SQTE (and standard deviation) at $\tau = 0.05$ from all models can be found in the Appendix.

\section{Conclusion and outlook}
We have proposed a flexible neural network-based causal spatial quantile regression framework to estimate heterogeneous treatment effects, which demonstrates high predictive power and is capable of identifying spatial quantile treatment effects effectively. To be specific, our model can capture complicated (non-linear, high-dimensional, complex interactions between covariates) dependencies and heterogeneous (spatially varying, non-linear, quantile-dependent) treatment effects, which can be estimated with relatively low errors. 

In the spatial setting where there exist spatial covariates and spatial random effects, including spatial information (coordinates and/or more spatial features at different resolutions) can further reduce prediction errors. Spatial confounding adjustment works well when hidden confounding variables exist by reducing prediction errors for models with no spatial features.

Even though this framework can estimate treatment effects at different quantile levels, prediction errors are higher in the lower and upper tail regions as expected. There are different possibilities for addressing this issue in future research, including applying a semiparametric transformation to the response variable or directly modifying the spline basis functions themselves in order to obtain alternative splines that are more in line with extreme-value theory. A relevant work that also addresses this problem can be found in \cite{majumder2025semi}.
Furthermore, recent work has been done to investigate the estimation and inference of extremal quantile treatment effects (i.e., for quantile levels outside of the range of the data) tailored for heavy-tailed distributions \citep{deuber2021estimation}, where they focus on extreme regions only. It would be interesting to investigate how to combine our causal quantile regression framework with theirs, to estimate spatial quantile treatment effects all the way from extremely low to extremely high quantiles, with a smooth transition in between.

We stress that there is more work to be done in future research to compare different spatial confounding models and adjustment methods with our approach. Moreover, while we have illustrated the methodology for a health-related application, the framework applies more generally. It would be interesting to use it in econometric applications, environmental risk attribution, and impact studies.

Finally, the doubly robust properties in the causal inference literature \citep{van2011targeted, kennedy2022semiparametric, kennedy2023towards} are particularly of interest as they offer two reliable pathways, propensity score regression and outcome regression, for consistent estimation. Our work utilizes flexible machine learning techniques to ensure consistent outcome regression while relaxing the stringent requirements typically placed on the propensity score model. Significant foundational theoretical work is necessary to support the proposed semiparametric neural-network-based quantile regression functional class, thereby enhancing and rigorously validating this comprehensive method’s robustness and efficacy.

\section*{Acknowledgments}
The first author was supported by funding from the King Abdullah University of Science and Technology (KAUST). This work was supported by grants from the Southeast National Synthesis Wildfire and the United States Geological Survey’s National Climate Adaptation Science Center (G21AC10045), the National Institutes of Health (R01ES031651-01), and the National Science Foundation (DMS2152887).

\baselineskip=12pt
\bibliographystyle{CUP}
\bibliography{Biblio}

\begin{thebibliography}{44}

\bibitem[Abrevaya \emph{et~al.}(2015)Abrevaya, Hsu and
  Lieli]{abrevaya2015estimating}
Abrevaya, J., Hsu, Y.-C. and Lieli, R.~P. (2015) Estimating conditional average
  treatment effects.
\newblock \emph{Journal of Business \& Economic Statistics} \textbf{33}(4),
  485--505.

\bibitem[Akbari \emph{et~al.}(2023)Akbari, Winter and Tomko]{akbari2023spatial}
Akbari, K., Winter, S. and Tomko, M. (2023) Spatial causality: A systematic
  review on spatial causal inference.
\newblock \emph{Geographical Analysis} \textbf{55}(1), 56--89.

\bibitem[Ando \emph{et~al.}(2023)Ando, Li and Lu]{ando2023spatial}
Ando, T., Li, K. and Lu, L. (2023) A spatial panel quantile model with
  unobserved heterogeneity.
\newblock \emph{Journal of Econometrics} \textbf{232}(1), 191--213.

\bibitem[Athey and Imbens(2015)]{athey2015machine}
Athey, S. and Imbens, G.~W. (2015) Machine learning methods for estimating
  heterogeneous causal effects.
\newblock \emph{Stat} \textbf{1050}(5), 1--26.

\bibitem[Bernardi and Guidolin(2023)]{bernardi2023determinants}
Bernardi, M. and Guidolin, M. (2023) The determinants of Airbnb prices in New
  York City: a spatial quantile regression approach.
\newblock \emph{Journal of the Royal Statistical Society Series C: Applied
  Statistics} \textbf{72}(1), 104--143.

\bibitem[de~Bont \emph{et~al.}(2024)de~Bont, Krishna, Stafoggia, Banerjee,
  Dholakia, Garg, Ingole, Jaganathan, Kloog, Lane \emph{et~al.}]{de2024ambient}
de~Bont, J., Krishna, B., Stafoggia, M., Banerjee, T., Dholakia, H., Garg, A.,
  Ingole, V., Jaganathan, S., Kloog, I., Lane, K. \emph{et~al.} (2024) Ambient
  air pollution and daily mortality in ten cities of {India}: a causal
  modelling study.
\newblock \emph{The Lancet Planetary Health} \textbf{8}(7), e433--e440.

\bibitem[Castillo-Mateo \emph{et~al.}(2023)Castillo-Mateo, As{\'\i}n,
  Cebri{\'a}n, Gelfand and Abaurrea]{castillo2023spatial}
Castillo-Mateo, J., As{\'\i}n, J., Cebri{\'a}n, A.~C., Gelfand, A.~E. and
  Abaurrea, J. (2023) Spatial quantile autoregression for season within year
  daily maximum temperature data.
\newblock \emph{The Annals of Applied Statistics} \textbf{17}(3), 2305--2325.

\bibitem[Chen \emph{et~al.}(2024)Chen, Li, Reich and Sun]{chen2020deepkriging}
Chen, W., Li, Y., Reich, B.~J. and Sun, Y. (2024) {DeepKriging}: Spatially
  dependent deep neural networks for spatial prediction.
\newblock \emph{Statistica Sinica} \textbf{34}.

\bibitem[Chen and Tokdar(2021)]{chen2021joint}
Chen, X. and Tokdar, S.~T. (2021) Joint quantile regression for spatial data.
\newblock \emph{Journal of the Royal Statistical Society Series B: Statistical
  Methodology} \textbf{83}(4), 826--852.

\bibitem[Chernozhukov and Hansen(2005)]{chernozhukov2005iv}
Chernozhukov, V. and Hansen, C. (2005) An {IV} model of quantile treatment
  effects.
\newblock \emph{Econometrica} \textbf{73}(1), 245--261.

\bibitem[Chernozhukov and Hansen(2006)]{chernozhukov2006instrumental}
Chernozhukov, V. and Hansen, C. (2006) Instrumental quantile regression
  inference for structural and treatment effect models.
\newblock \emph{Journal of Econometrics} \textbf{132}(2), 491--525.

\bibitem[Davis \emph{et~al.}(2019)Davis, Neelon, Nietert, Hunt, Burgette,
  Lawson and Egede]{davis2019addressing}
Davis, M.~L., Neelon, B., Nietert, P.~J., Hunt, K.~J., Burgette, L.~F., Lawson,
  A.~B. and Egede, L.~E. (2019) Addressing geographic confounding through
  spatial propensity scores: a study of racial disparities in diabetes.
\newblock \emph{Statistical Methods in Medical Research} \textbf{28}(3),
  734--748.

\bibitem[Deuber \emph{et~al.}(2024)Deuber, Li, Engelke and
  Maathuis]{deuber2021estimation}
Deuber, D., Li, J., Engelke, S. and Maathuis, M.~H. (2024) Estimation and
  inference of extremal quantile treatment effects for heavy-tailed
  distributions.
\newblock \emph{Journal of the American Statistical Association}
  \textbf{119}(547), 2206--2216.

\bibitem[Firpo(2007)]{firpo2007efficient}
Firpo, S. (2007) Efficient semiparametric estimation of quantile treatment
  effects.
\newblock \emph{Econometrica} \textbf{75}(1), 259--276.

\bibitem[Gao \emph{et~al.}(2022)Gao, Wang, Stein and Chen]{gao2022causal}
Gao, B., Wang, J., Stein, A. and Chen, Z. (2022) Causal inference in spatial
  statistics.
\newblock \emph{Spatial statistics} \textbf{50}, 100621.

\bibitem[Giffin \emph{et~al.}(2023)Giffin, Reich, Yang and
  Rappold]{giffin2023generalized}
Giffin, A., Reich, B., Yang, S. and Rappold, A. (2023) Generalized propensity
  score approach to causal inference with spatial interference.
\newblock \emph{Biometrics} \textbf{79}(3), 2220--2231.

\bibitem[Gilbert \emph{et~al.}(2021)Gilbert, Datta, Casey and
  Ogburn]{gilbert2021causal}
Gilbert, B., Datta, A., Casey, J.~A. and Ogburn, E.~L. (2021) A causal
  inference framework for spatial confounding.
\newblock \emph{arXiv preprint arXiv:2112.14946} .

\bibitem[Hitsch \emph{et~al.}(2024)Hitsch, Misra and
  Zhang]{hitsch2024heterogeneous}
Hitsch, G.~J., Misra, S. and Zhang, W.~W. (2024) Heterogeneous treatment
  effects and optimal targeting policy evaluation.
\newblock \emph{Quantitative Marketing and Economics} \textbf{22}(2), 115--168.

\bibitem[Jarner \emph{et~al.}(2002)Jarner, Diggle and
  Chetwynd]{jarner2002estimation}
Jarner, M.~F., Diggle, P. and Chetwynd, A.~G. (2002) Estimation of spatial
  variation in risk using matched case-control data.
\newblock \emph{Biometrical Journal: Journal of Mathematical Methods in
  Biosciences} \textbf{44}(8), 936--945.

\bibitem[Kennedy(2023)]{kennedy2023towards}
Kennedy, E.~H. (2023) Towards optimal doubly robust estimation of heterogeneous
  causal effects.
\newblock \emph{Electronic Journal of Statistics} \textbf{17}(2), 3008--3049.

\bibitem[Kennedy(2024)]{kennedy2022semiparametric}
Kennedy, E.~H. (2024) Semiparametric doubly robust targeted double machine
  learning: a review.
\newblock In \emph{Handbook of Statistical Methods for Precision Medicine}.
  Chapman \& Hall/CRC.
\newblock Editors E. Laber, B. Chakraborty, E. E. M. Moodie, T. Cai, and M. van
  der Laan.

\bibitem[Kent \emph{et~al.}(2018)Kent, Steyerberg and
  Van~Klaveren]{kent2018personalized}
Kent, D.~M., Steyerberg, E. and Van~Klaveren, D. (2018) Personalized evidence
  based medicine: predictive approaches to heterogeneous treatment effects.
\newblock \emph{BMJ} \textbf{363}.

\bibitem[Kim \emph{et~al.}(2025)Kim, Wang and Wang]{kim2025estimation}
Kim, M., Wang, L. and Wang, H.~J. (2025) Estimation and Inference of Quantile
  Spatially Varying Coefficient Models Over Complicated Domains.
\newblock \emph{Journal of the American Statistical Association}
  (just-accepted), 1--23.

\bibitem[Van~der Laan \emph{et~al.}(2011)Van~der Laan, Rose
  \emph{et~al.}]{van2011targeted}
Van~der Laan, M.~J., Rose, S. \emph{et~al.} (2011) \emph{Targeted Learning:
  Causal Inference for Observational and Experimental Data}.
\newblock Springer.

\bibitem[Majumder and Richards(2025)]{majumder2025semi}
Majumder, R. and Richards, J. (2025) Semi-parametric bulk and tail regression
  using spline-based neural networks.
\newblock \emph{arXiv preprint arXiv:2504.19994} .

\bibitem[Neyman(1923)]{neyman1923applications}
Neyman, J. (1923) Sur les applications de la th{\'e}orie des probabilit{\'e}s
  aux experiences agricoles: Essai des principes.
\newblock \emph{Roczniki Nauk Rolniczych} \textbf{10}(1), 1--51.

\bibitem[Nychka \emph{et~al.}(2015)Nychka, Bandyopadhyay, Hammerling, Lindgren
  and Sain]{nychka2015multiresolution}
Nychka, D., Bandyopadhyay, S., Hammerling, D., Lindgren, F. and Sain, S. (2015)
  A multiresolution {Gaussian} process model for the analysis of large spatial
  datasets.
\newblock \emph{Journal of Computational and Graphical Statistics}
  \textbf{24}(2), 579--599.

\bibitem[Papadogeorgou and Samanta(2023)]{papadogeorgou2023spatial}
Papadogeorgou, G. and Samanta, S. (2023) Spatial causal inference in the
  presence of unmeasured confounding and interference.
\newblock \emph{arXiv preprint arXiv:2303.08218} .

\bibitem[Reich(2012)]{reich2012spatiotemporal}
Reich, B.~J. (2012) Spatiotemporal quantile regression for detecting
  distributional changes in environmental processes.
\newblock \emph{Journal of the Royal Statistical Society Series C: Applied
  Statistics} \textbf{61}(4), 535--553.

\bibitem[Reich \emph{et~al.}(2011)Reich, Fuentes and Dunson]{reich2011bayesian}
Reich, B.~J., Fuentes, M. and Dunson, D.~B. (2011) Bayesian spatial quantile
  regression.
\newblock \emph{Journal of the American Statistical Association}
  \textbf{106}(493), 6--20.

\bibitem[Reich \emph{et~al.}(2021)Reich, Yang, Guan, Giffin, Miller and
  Rappold]{reich2021review}
Reich, B.~J., Yang, S., Guan, Y., Giffin, A.~B., Miller, M.~J. and Rappold, A.
  (2021) A review of spatial causal inference methods for environmental and
  epidemiological applications.
\newblock \emph{International Statistical Review} \textbf{89}(3), 605--634.

\bibitem[Rockl{\"o}v \emph{et~al.}(2023)Rockl{\"o}v, Semenza, Dasgupta,
  Robinson, Abd El~Wahed, Alcayna, Arn{\'e}s-Sanz, Bailey, B{\"a}rnighausen,
  Bartumeus \emph{et~al.}]{rocklov2023decision}
Rockl{\"o}v, J., Semenza, J.~C., Dasgupta, S., Robinson, E.~J., Abd El~Wahed,
  A., Alcayna, T., Arn{\'e}s-Sanz, C., Bailey, M., B{\"a}rnighausen, T.,
  Bartumeus, F. \emph{et~al.} (2023) Decision-support tools to build climate
  resilience against emerging infectious diseases in {Europe} and beyond.
\newblock \emph{The Lancet Regional Health--Europe} \textbf{32}.

\bibitem[Rubin(1974)]{rubin1974estimating}
Rubin, D.~B. (1974) Estimating causal effects of treatments in randomized and
  nonrandomized studies.
\newblock \emph{Journal of Educational Psychology} \textbf{66}(5), 688.

\bibitem[Schnell and Papadogeorgou(2020)]{schnell2020mitigating}
Schnell, P.~M. and Papadogeorgou, G. (2020) Mitigating unobserved spatial
  confounding when estimating the effect of supermarket access on
  cardiovascular disease deaths.
\newblock \emph{Annals of Applied Statistics} \textbf{14}(4).

\bibitem[Sun \emph{et~al.}(2021)Sun, Moodie and
  Ne{{s}}lehov{\'a}]{sun2021causal}
Sun, S., Moodie, E.~E. and Ne{{s}}lehov{\'a}, J.~G. (2021) Causal inference for
  quantile treatment effects.
\newblock \emph{Environmetrics} \textbf{32}(4), e2668.

\bibitem[Venkatasubramaniam \emph{et~al.}(2023)Venkatasubramaniam, Mateen,
  Shields, Hattersley, Jones, Vollmer and
  Dennis]{venkatasubramaniam2023comparison}
Venkatasubramaniam, A., Mateen, B.~A., Shields, B.~M., Hattersley, A.~T.,
  Jones, A.~G., Vollmer, S.~J. and Dennis, J.~M. (2023) Comparison of causal
  forest and regression-based approaches to evaluate treatment effect
  heterogeneity: an application for type 2 diabetes precision medicine.
\newblock \emph{BMC Medical Informatics and Decision Making} \textbf{23}(1),
  110.

\bibitem[Wager and Athey(2018)]{wager2018estimation}
Wager, S. and Athey, S. (2018) Estimation and inference of heterogeneous
  treatment effects using random forests.
\newblock \emph{Journal of the American Statistical Association}
  \textbf{113}(523), 1228--1242.

\bibitem[W{\"u}thrich(2019)]{wuthrich2019closed}
W{\"u}thrich, K. (2019) A closed-form estimator for quantile treatment effects
  with endogeneity.
\newblock \emph{Journal of Econometrics} \textbf{210}(2), 219--235.

\bibitem[Xu \emph{et~al.}(2018)Xu, Daniels and Winterstein]{xu2018bayesian}
Xu, D., Daniels, M.~J. and Winterstein, A.~G. (2018) A {Bayesian} nonparametric
  approach to causal inference on quantiles.
\newblock \emph{Biometrics} \textbf{74}(3), 986--996.

\bibitem[Xu \emph{et~al.}(2022{a})Xu, Majumder and Reich]{xu2022spqr}
Xu, S.~G., Majumder, R. and Reich, B.~J. (2022{a}) {SPQR}: An {R} Package for
  Semi-Parametric Density and Quantile Regression.
\newblock \emph{arXiv preprint arXiv:2210.14482} .

\bibitem[Xu and Reich(2023)]{xu2021bayesian}
Xu, S.~G. and Reich, B.~J. (2023) Bayesian nonparametric quantile process
  regression and estimation of marginal quantile effects.
\newblock \emph{Biometrics} \textbf{79}, 151--164.

\bibitem[Xu \emph{et~al.}(2022{b})Xu, Yang and Reich]{xu2022bayesian}
Xu, S.~G., Yang, S. and Reich, B.~J. (2022{b}) A Bayesian Semiparametric Method
  For Estimating Causal Quantile Effects.
\newblock \emph{arXiv preprint arXiv:2211.01591} .

\bibitem[Yang \emph{et~al.}(2024)Yang, Eckles, Dhillon and
  Aral]{yang2024targeting}
Yang, J., Eckles, D., Dhillon, P. and Aral, S. (2024) Targeting for long-term
  outcomes.
\newblock \emph{Management Science} \textbf{70}(6), 3841--3855.

\bibitem[Zorzetto \emph{et~al.}(2024)Zorzetto, Bargagli-Stoffi, Canale and
  Dominici]{zorzetto2024confounder}
Zorzetto, D., Bargagli-Stoffi, F.~J., Canale, A. and Dominici, F. (2024)
  Confounder-dependent {Bayesian} mixture model: Characterizing heterogeneity
  of causal effects in air pollution epidemiology.
\newblock \emph{Biometrics} \textbf{80}(2), ujae025.

\end{thebibliography}

\clearpage
\appendix
\counterwithin{figure}{section}

\section{Additional simulation results}\label{A:sd_simu}
\subsection{Scenario 1: Unconfoundedness}

\begin{figure}[H]
\centering
\includegraphics[width=1\linewidth]{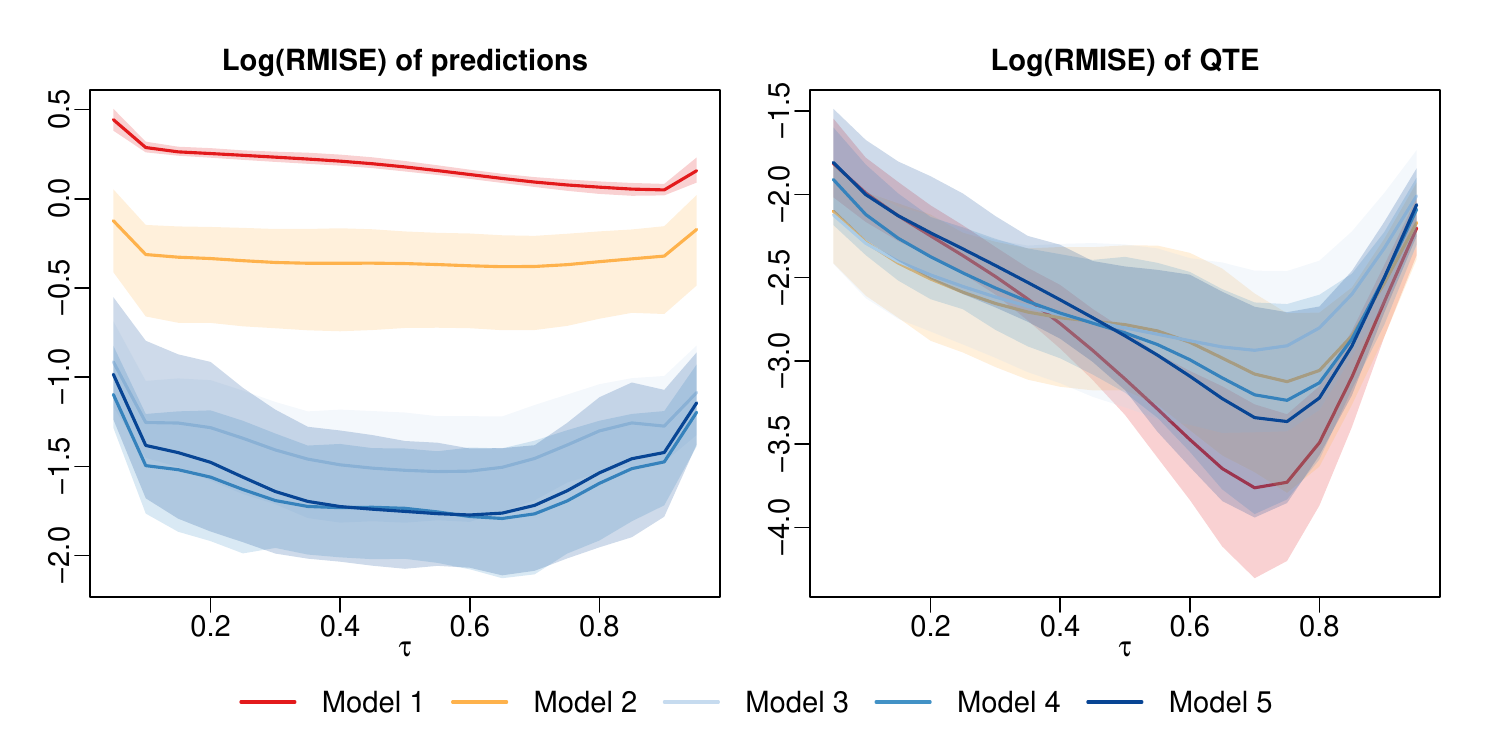}
\vspace{-1.2cm}
\caption{[Scenario 1: Unconfoundedness.] Log(RMISE) of predicted responses (left panel) and QTE estimates (right panel) at different quantile levels from five models. The shaded areas correspond to 95\% confidence intervals from 100 simulations.}\label{fig:p0_rmise}
\end{figure}

\begin{figure}[H]
\centering
\includegraphics[width=1\linewidth]{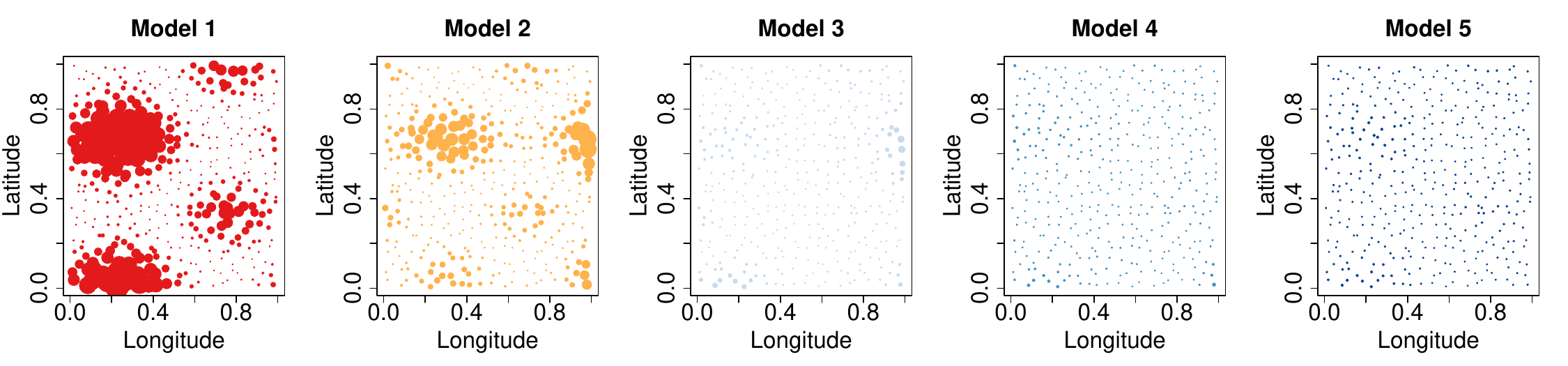}
\caption{[Scenario 1: Unconfoundedness.] Spatial patterns of the mean (of 100 simulation results) RMISE for predicting the response at the quantile level $\tau = 0.05$ from the five models. The size of the points is proportional to the RMISE values.}	\label{fig:p0_errmat}
\end{figure}

\begin{figure}[H]
\centering
\includegraphics[width=1\linewidth]{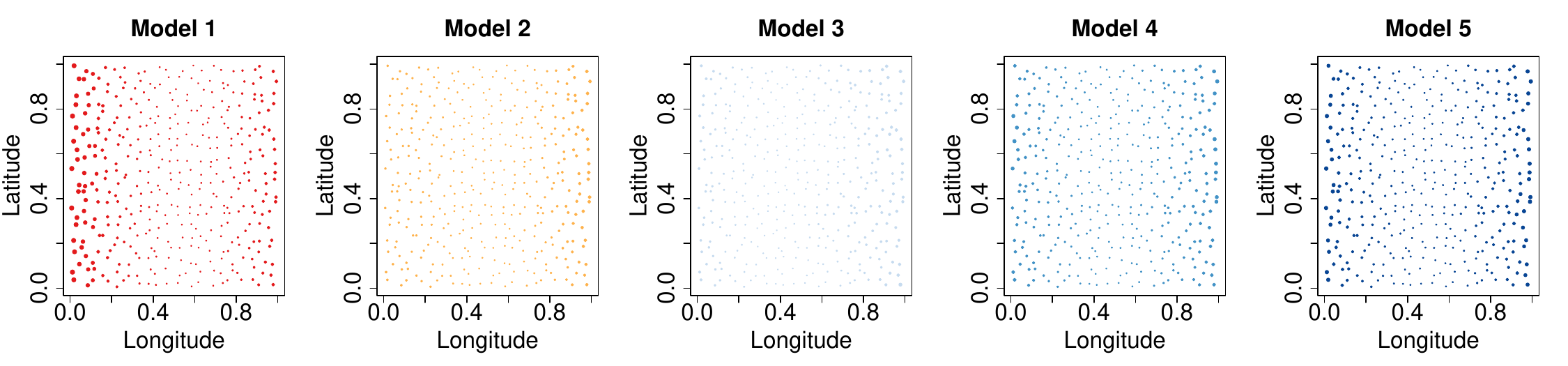}
\caption{[Scenario 1: Unconfoundedness.] Spatial patterns of the mean (of 100 simulation results) RMISE for estimating SQTE at the quantile level $\tau = 0.05$ from the five models. The size of the points is proportional to the RMISE values.}	\label{fig:p0_qfmat}
\end{figure}

\begin{figure}[H]
\centering
\includegraphics[width=1\linewidth]{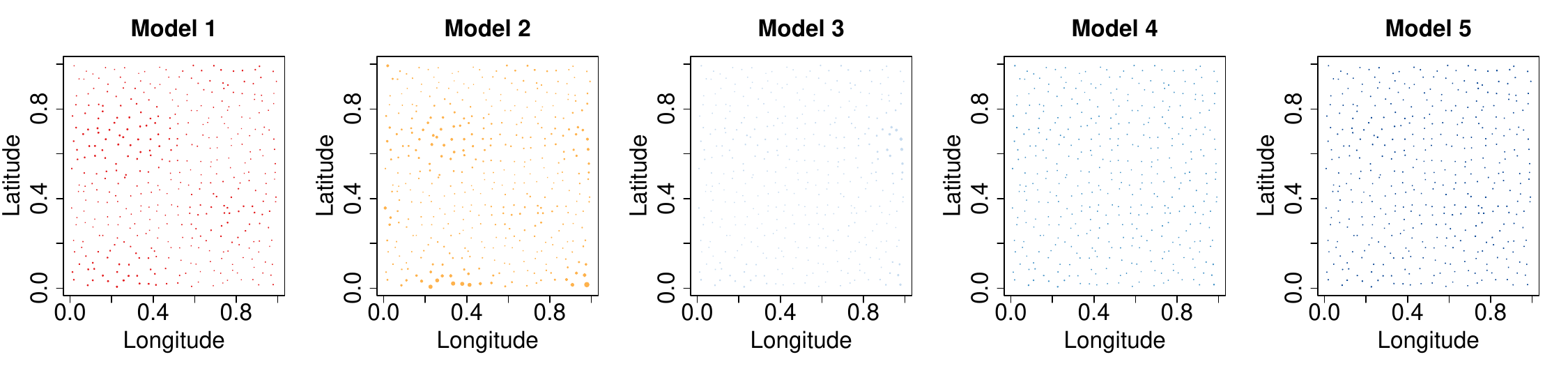}
\caption{[Scenario 1: Unconfoundedness.] Spatial patterns of the standard deviation (of 100 simulation results) RMISE for predicting the response at the quantile level $\tau = 0.05$ from the five models. The size of the points is proportional to the RMISE values.}	
\end{figure}

\begin{figure}[H]
\centering
\includegraphics[width=1\linewidth]{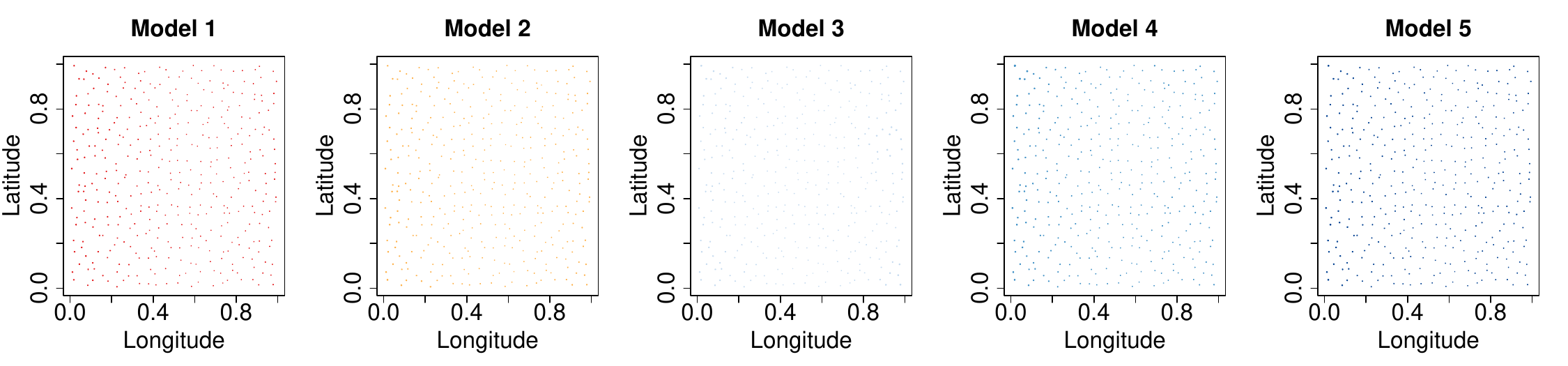}
\caption{[Scenario 1: Unconfoundedness.] Spatial patterns of the standard deviation (of 100 simulation results) RMISE for estimating SQTE at the quantile level $\tau = 0.05$ from the five models. The size of the points is proportional to the RMISE values.}
\end{figure}

\newpage
\subsection{Scenario 2: With observed confounders}
\begin{figure}[H]
\centering
\includegraphics[width=1\linewidth]{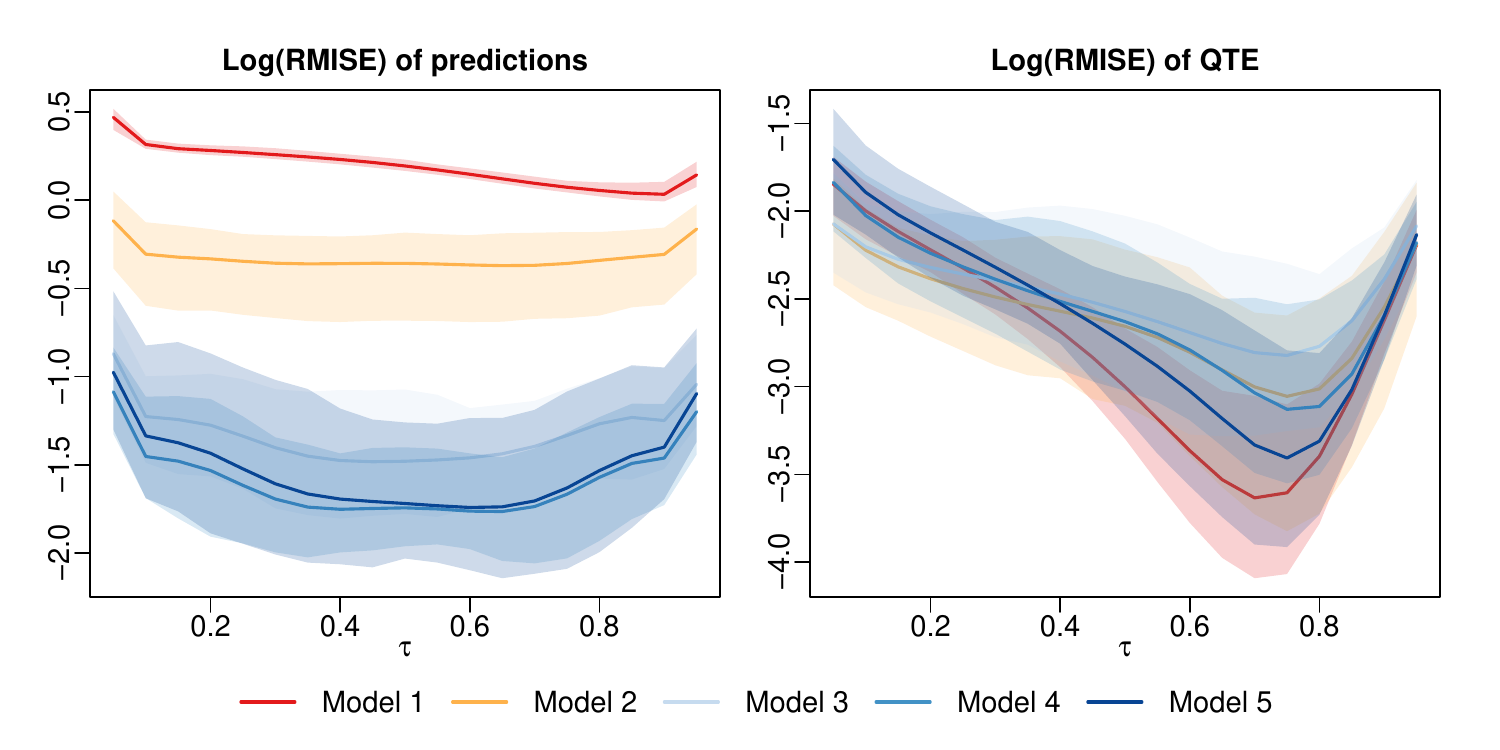}
\vspace{-1.1cm}
\caption{[Scenario 2: With observed confounders.] Log(RMISE) of predicted responses (left panel) and QTE estimates (right panel) at different quantile levels from five models. The shaded areas correspond to 95\% confidence intervals from 100 simulations.}\label{fig:p1_rmise}
\end{figure}

\begin{figure}[H]
\centering
\includegraphics[width=1\linewidth]{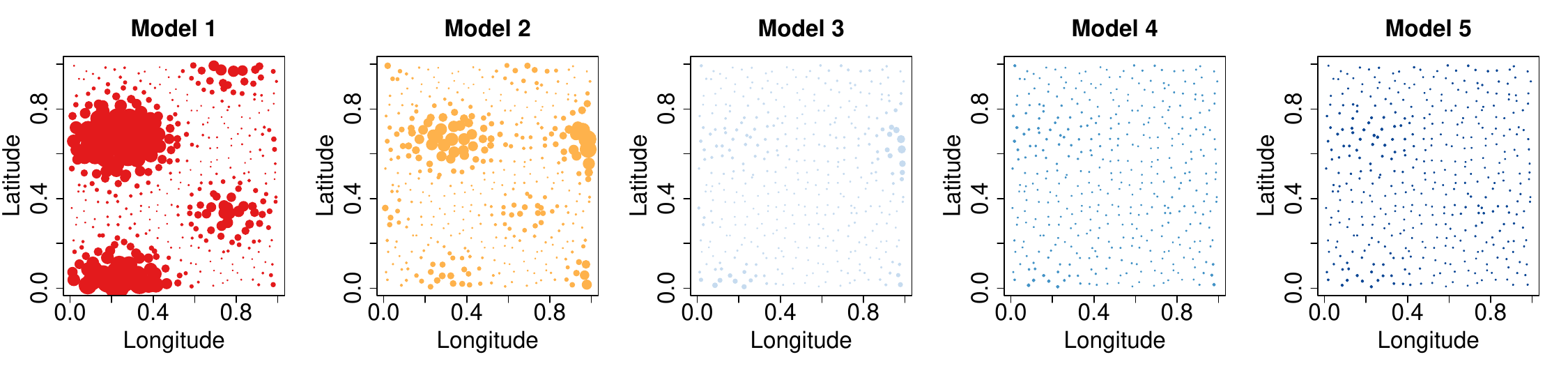}
\caption{[Scenario 2: With observed confounders.] Spatial patterns of the mean (of 100 simulation results) RMISE for predicting the response at the quantile level $\tau = 0.05$ from the five models. The size of the points is proportional to the RMISE values.}	\label{fig:p1_errmat}
\end{figure}

\begin{figure}[H]
\centering
\includegraphics[width=1\linewidth]{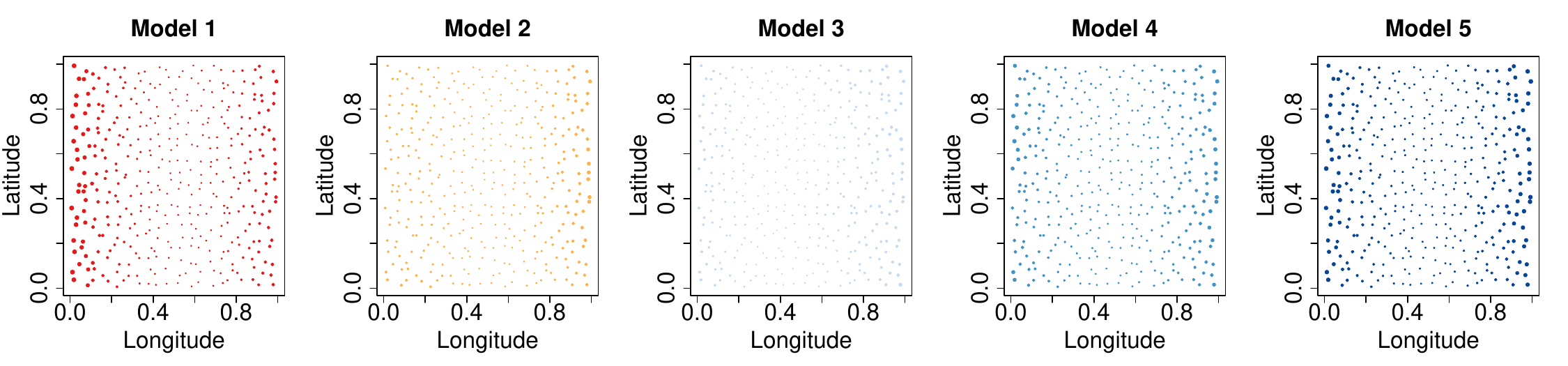}
\caption{[Scenario 2: With observed confounders.] Spatial patterns of the mean (of 100 simulation results) RMISE for estimating SQTE at the quantile level $\tau = 0.05$ from the five models. The size of the points is proportional to the RMISE values.}	\label{fig:p1_qfmat}
\end{figure}

\begin{figure}[H]
\centering
\includegraphics[width=1\linewidth]{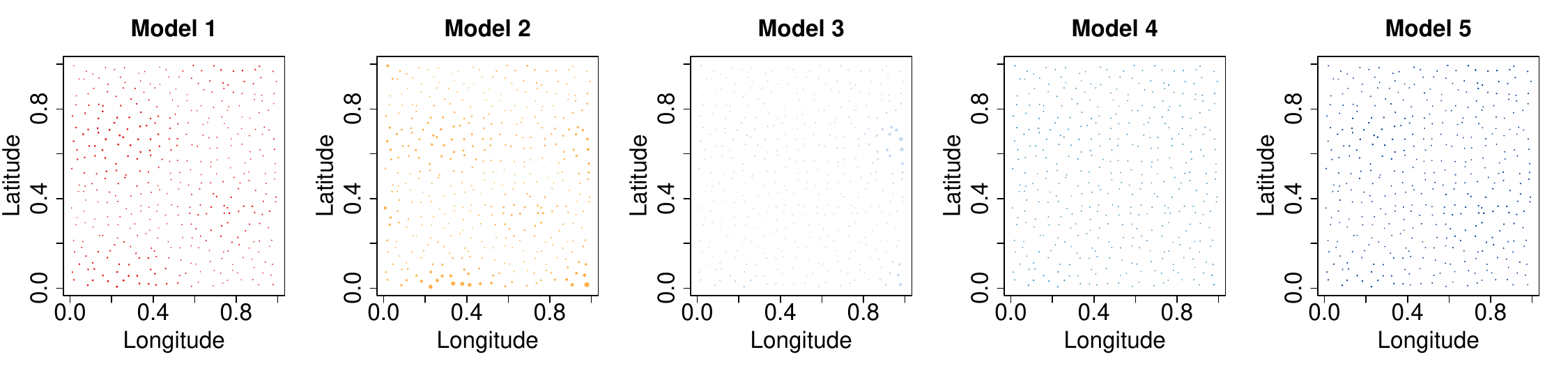}
\caption{[Scenario 2: With observed confounders.] Spatial patterns of the standard deviation (of 100 simulation results) RMISE for predicting the response at the quantile level $\tau = 0.05$ from the five models. The size of the points is proportional to the RMISE values.}	
\end{figure}

\begin{figure}[H]
\centering
\includegraphics[width=1\linewidth]{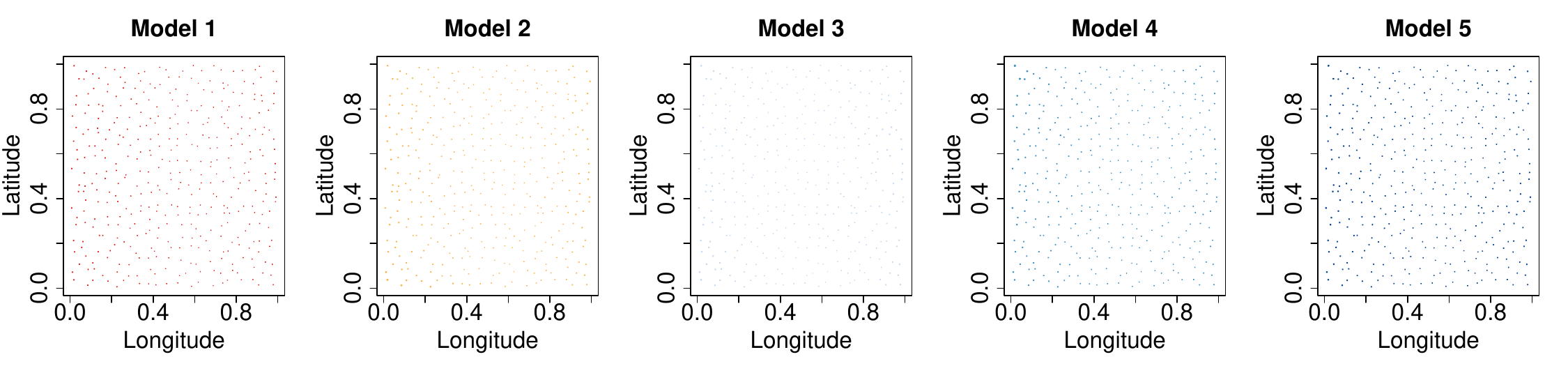}
\caption{[Scenario 2: With observed confounders.] Spatial patterns of the standard deviation (of 100 simulation results) RMISE for estimating SQTE at the quantile level $\tau = 0.05$ from the five models. The size of the points is proportional to the RMISE values.}
\end{figure}

\newpage
\subsection{Scenario 3: With hidden confounders}
\begin{figure}[H]
\centering
\includegraphics[width=1\linewidth]{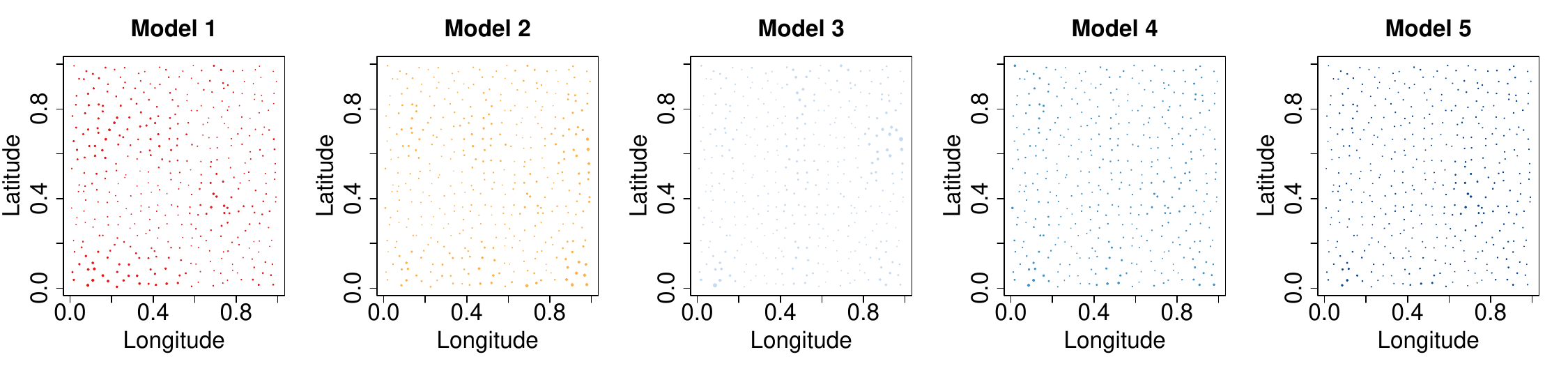}
\includegraphics[width=1\linewidth]{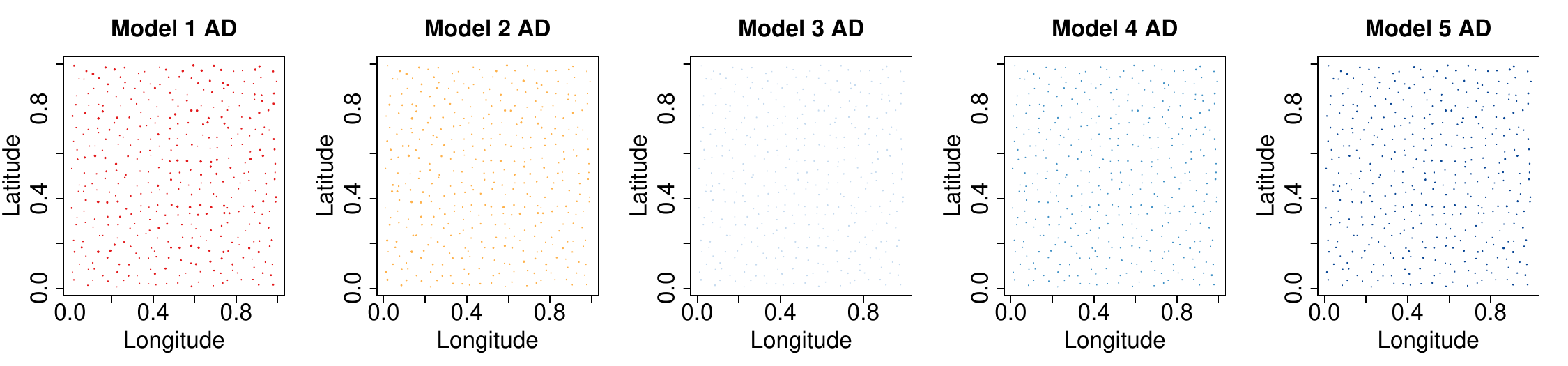}
\caption{[Scenario 3: With hidden confounders.] Spatial patterns of the standard deviation (of 100 simulation results) RMISE for predicting the response at the quantile level $\tau = 0.05$ from the five models before (first row) and after (second row) spatial confounding adjustment. The size of the points is proportional to the RMISE values.}
\end{figure}

\begin{figure}[H]
\centering
\includegraphics[width=1\linewidth]{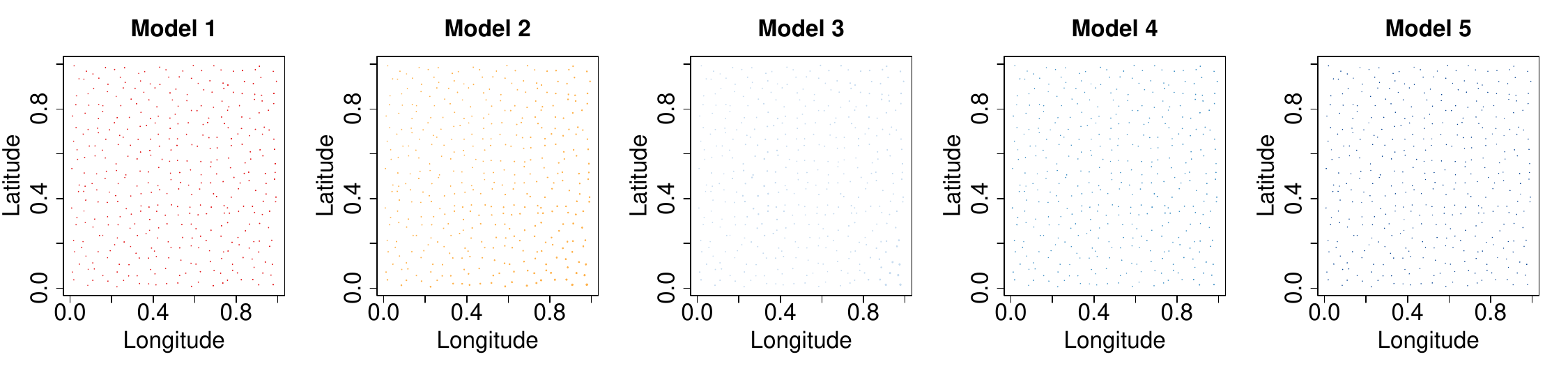}
\includegraphics[width=1\linewidth]{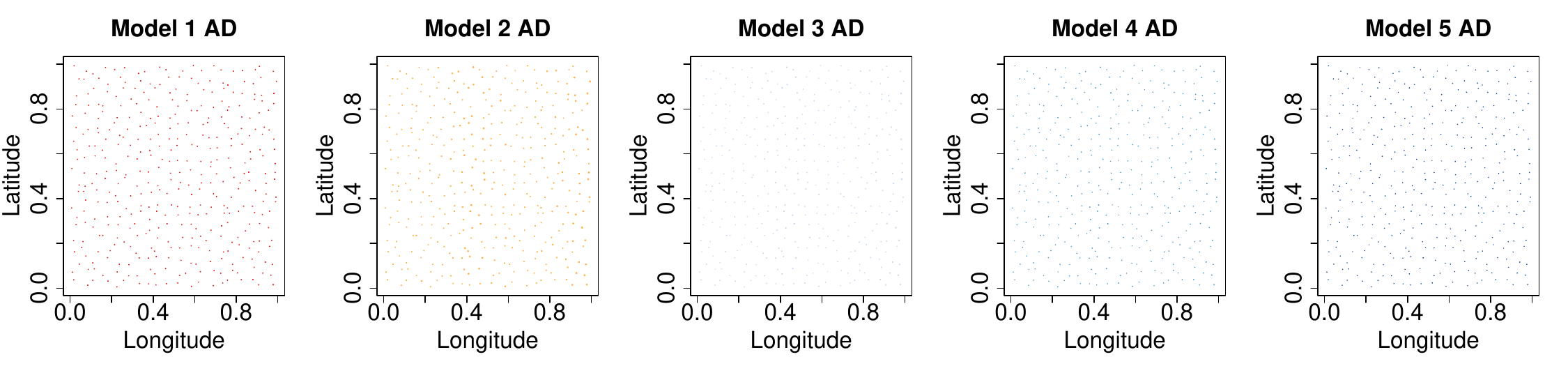}
\caption{[Scenario 3: With hidden confounders.] Spatial patterns of the standard deviation (of 100 simulation results) RMISE for estimating SQTE at the quantile level $\tau = 0.05$ from the five models before (first row) and after (second row) spatial confounding adjustment. The size of the points is proportional to the RMISE values.}
\end{figure}

\section{Additional application results}
\subsection{Sensitivity analysis for spatial confounding adjustment}\label{A:sensitivity}
\begin{figure}[H]
	\centering
  \includegraphics[width=1\linewidth]{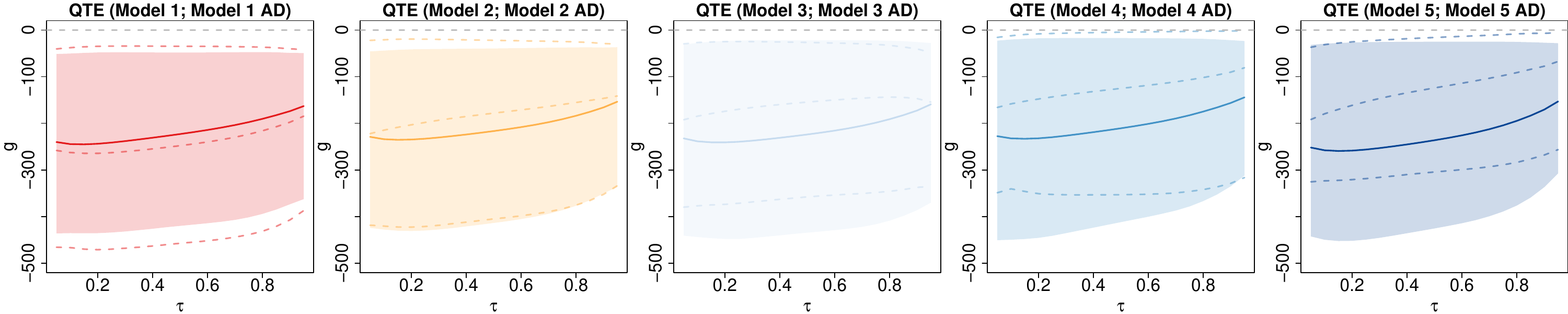}
  \includegraphics[width=1\linewidth]{app_qte_new202412_both1.pdf}
  \includegraphics[width=1\linewidth]{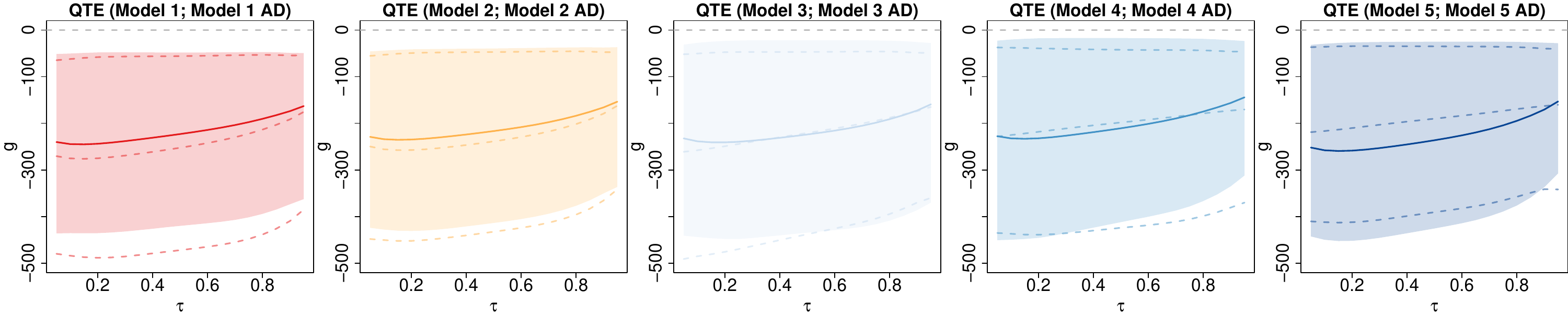}
   \label{fig:distance}
	\caption{Sensitiviety analysis for spatial adjustment, with 10\% (first row), 20\% (second row), 50\% (the third row) neighboring observations included, where we display the estimated QTE (in grams) and 95\% confidence intervals from the five models before (solid lines and shaded areas) and after (dashed lines) spatial confounding adjustment at different quantile levels, averaged over all the spatial locations.}
\end{figure}

\subsection{Estimated SQTE and standard deviation at the quantile level \\$\tau = 0.05$ for all models}\label{A:app_all_models}
\begin{figure}[H]
	\centering
 \includegraphics[width=1\linewidth]{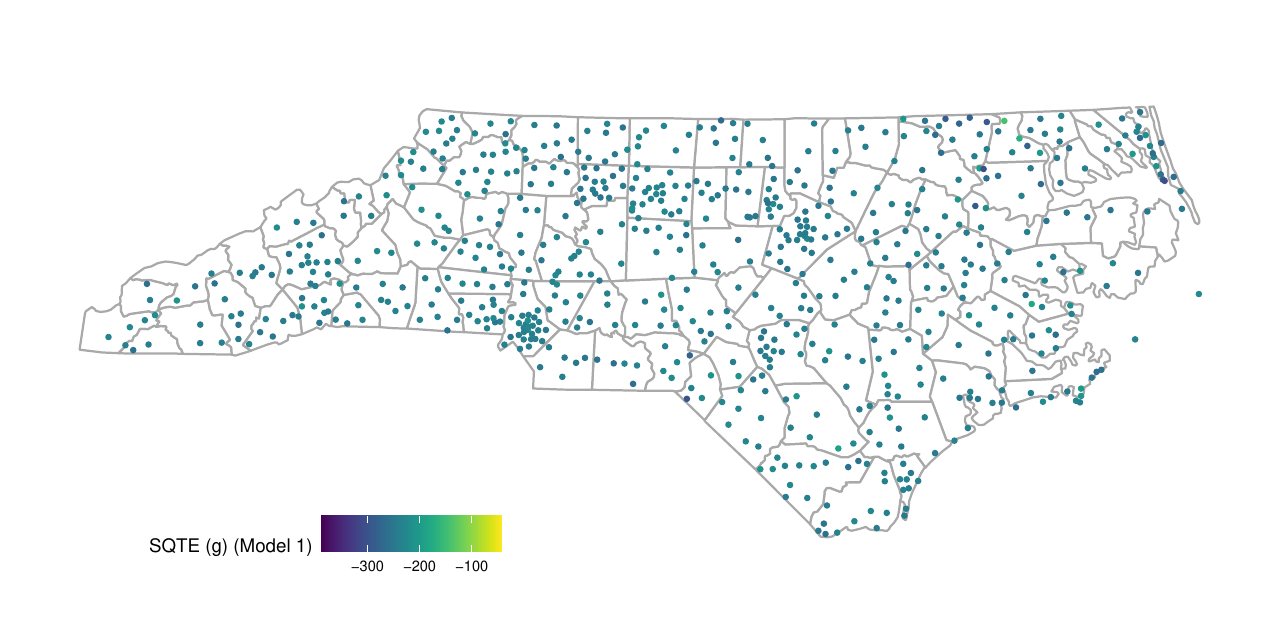}
  \includegraphics[width=1\linewidth]{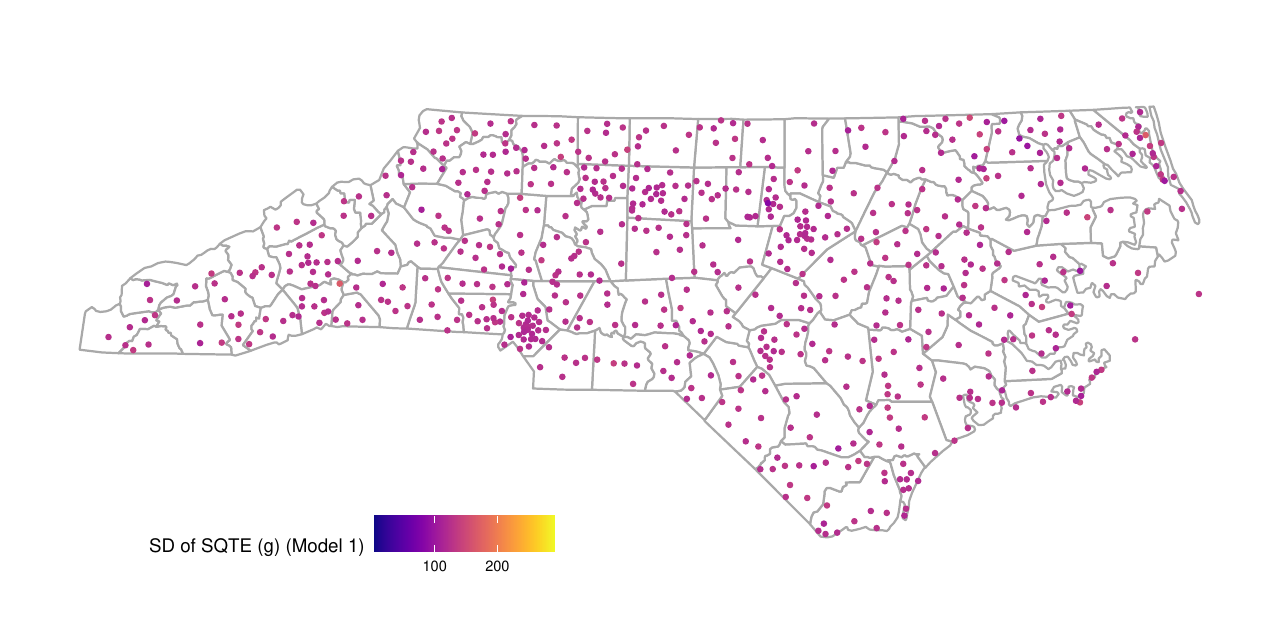}
  \caption{Estimated SQTE and standard deviation (SD) (in grams) of maternal smoking on the birth weight of newborn babies at $\tau = 0.05$ from Model 1 at all locations in North Carolina.}
\end{figure}
\begin{figure}[H]
\centering
\includegraphics[width=1\linewidth]{app001new_map_m1_mean_match.pdf}
  \includegraphics[width=1\linewidth]{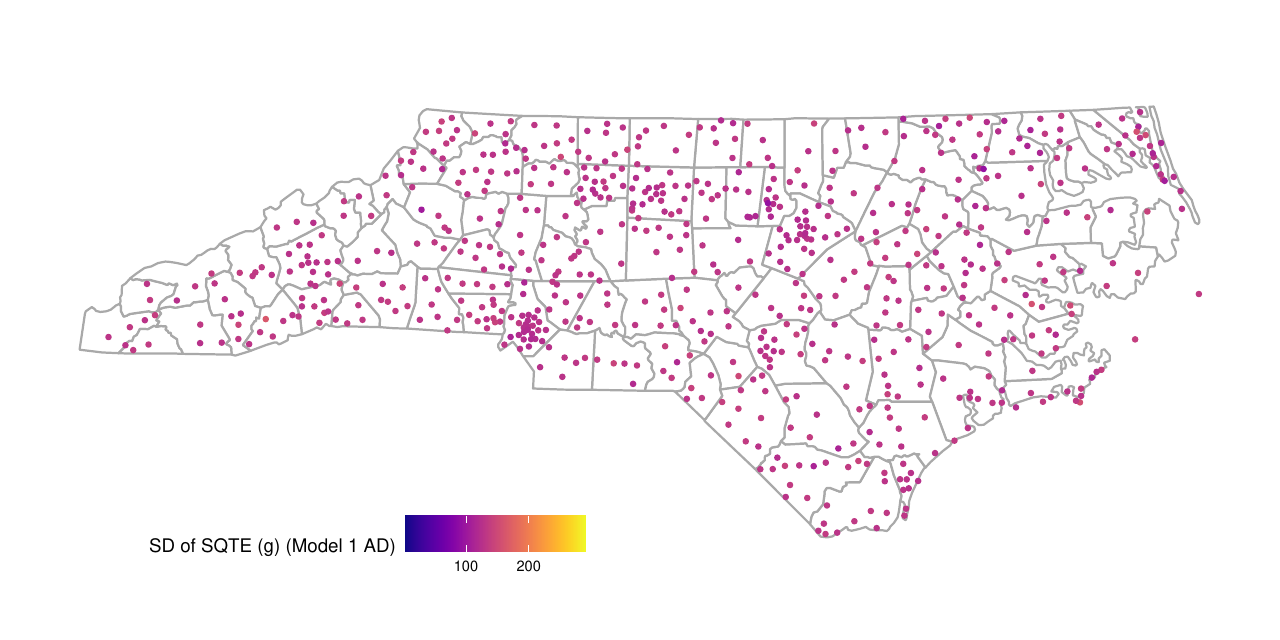}
    \caption{Estimated SQTE and standard deviation (SD) (in grams) of maternal smoking on the birth weight of newborn babies at $\tau = 0.05$ from Model 1 AD at all locations in North Carolina.}
\end{figure}

\begin{figure}[H]
	\centering
 \includegraphics[width=1\linewidth]{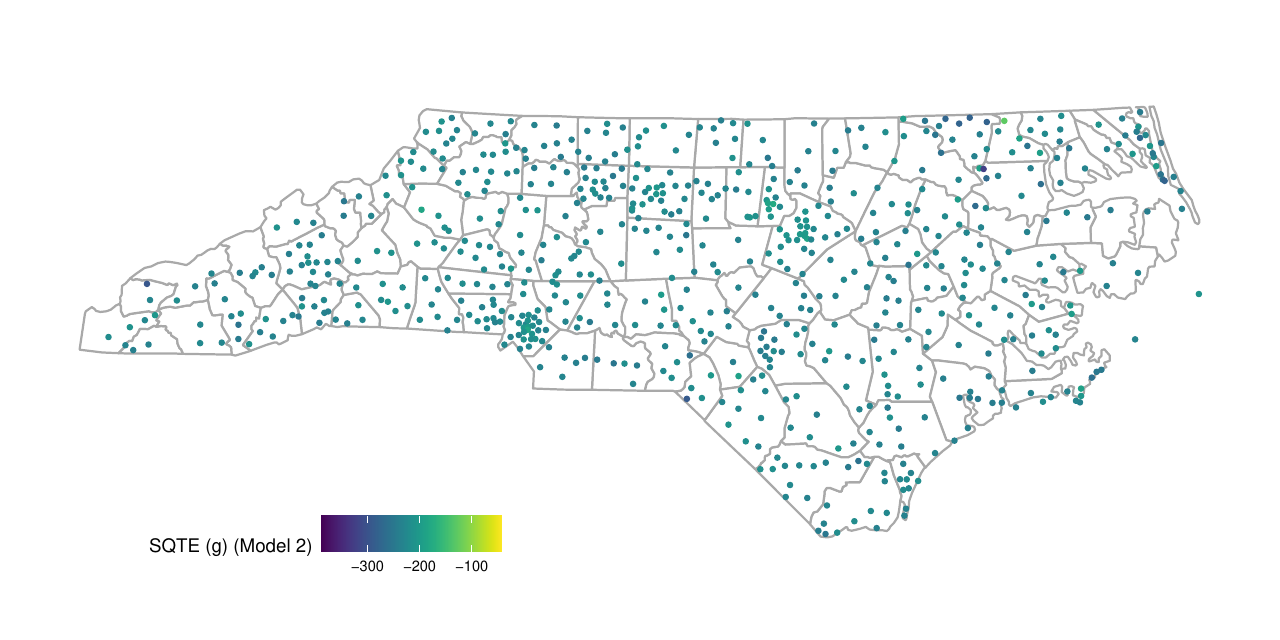}
  \includegraphics[width=1\linewidth]{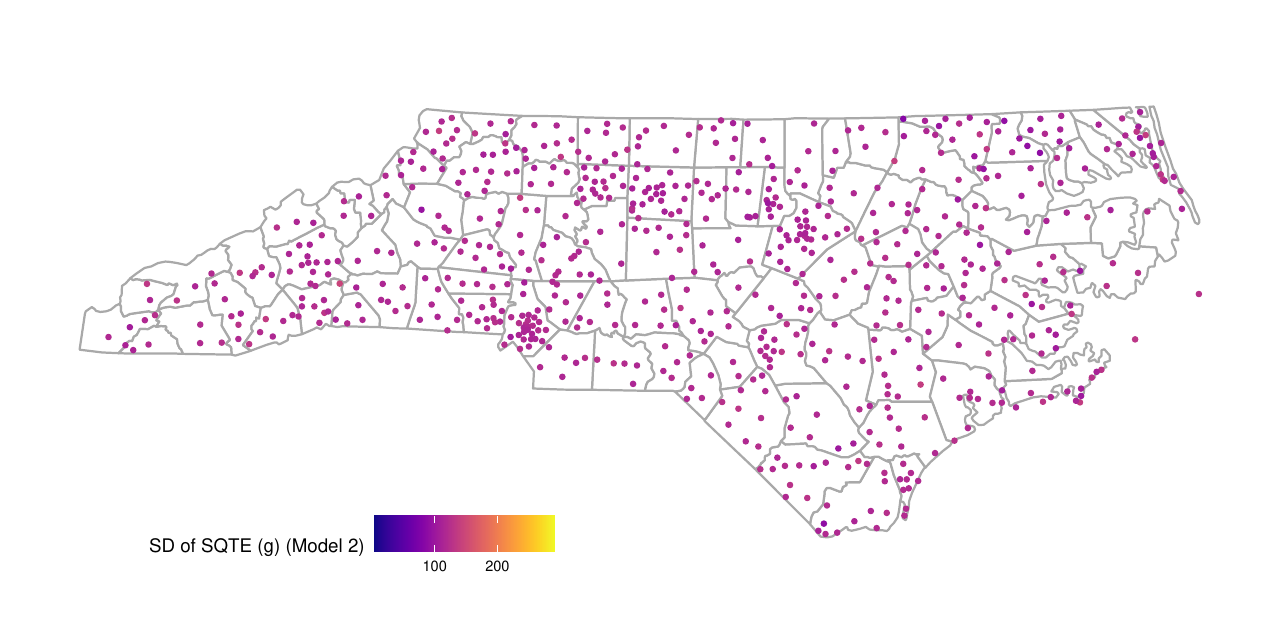}
    \caption{Estimated SQTE and standard deviation (SD) (in grams) of maternal smoking on the birth weight of newborn babies at $\tau = 0.05$ from Model 2 at all locations in North Carolina.}
\end{figure}

\begin{figure}[H]
\centering
\includegraphics[width=1\linewidth]{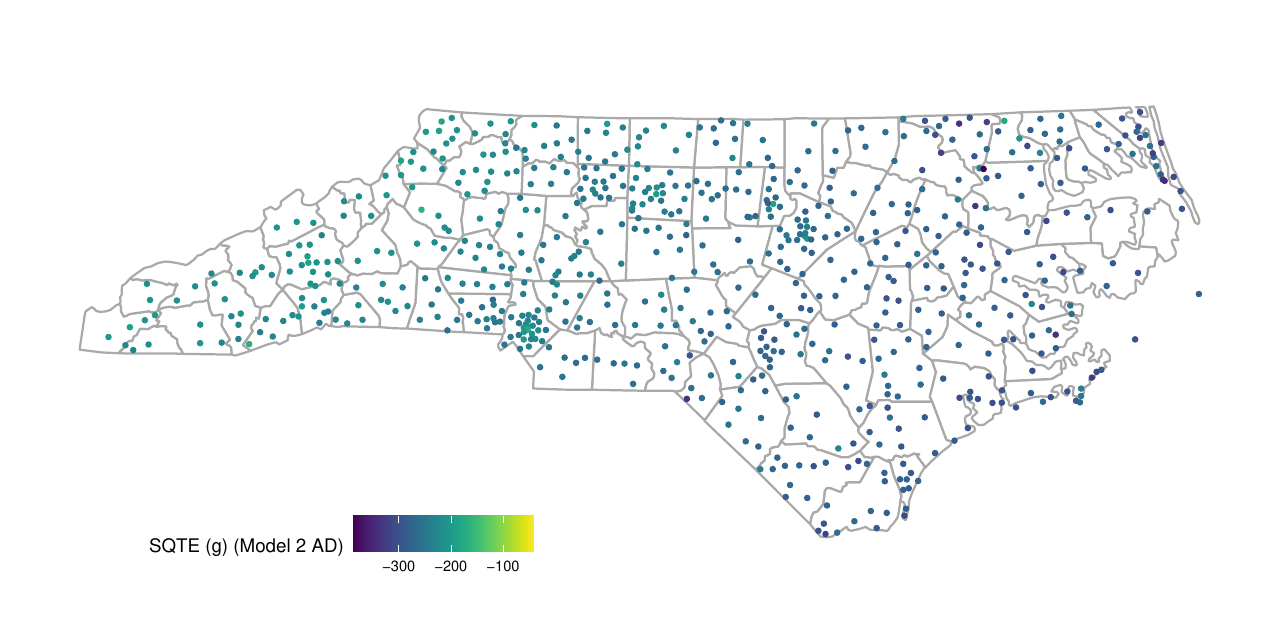}
  \includegraphics[width=1\linewidth]{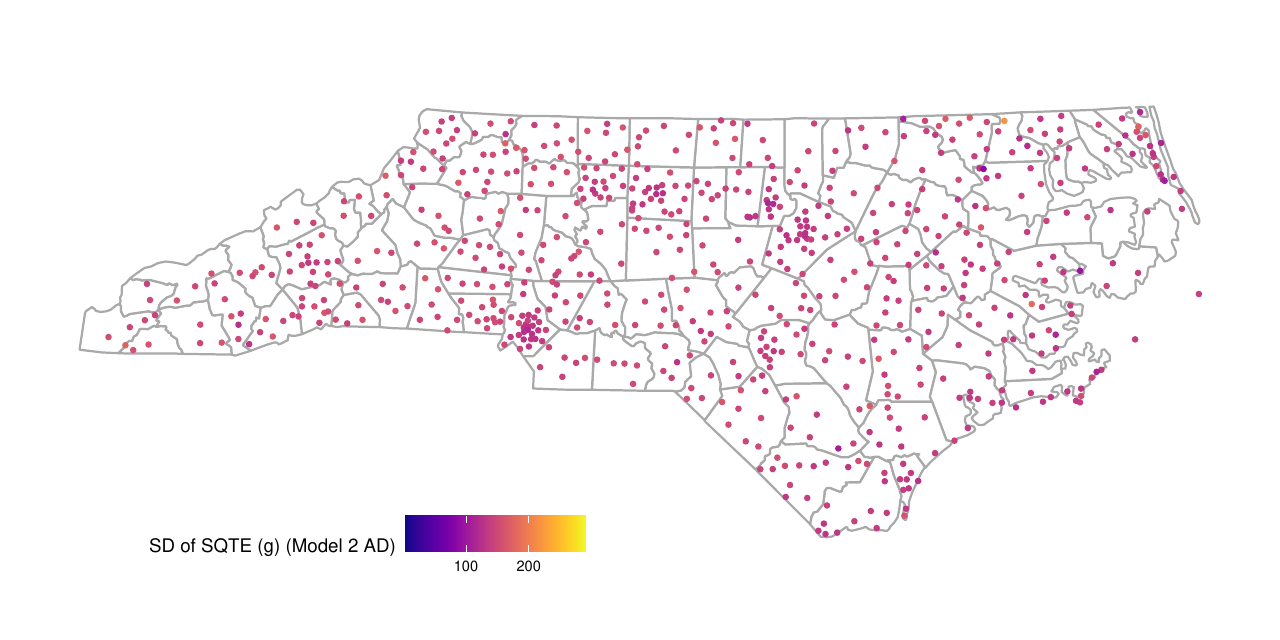}
    \caption{Estimated SQTE and standard deviation (SD) (in grams) of maternal smoking on the birth weight of newborn babies at $\tau = 0.05$ from Model 2 AD at all locations in North Carolina.}
\end{figure}

\begin{figure}[H]
	\centering
 \includegraphics[width=1\linewidth]{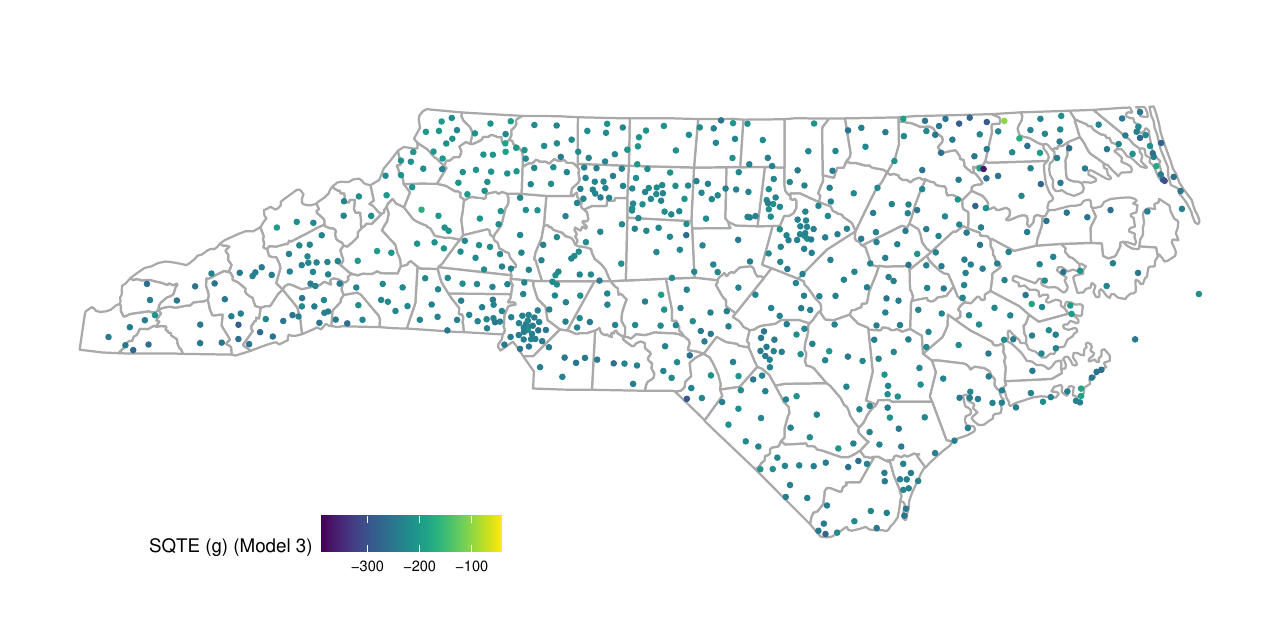}
  \includegraphics[width=1\linewidth]{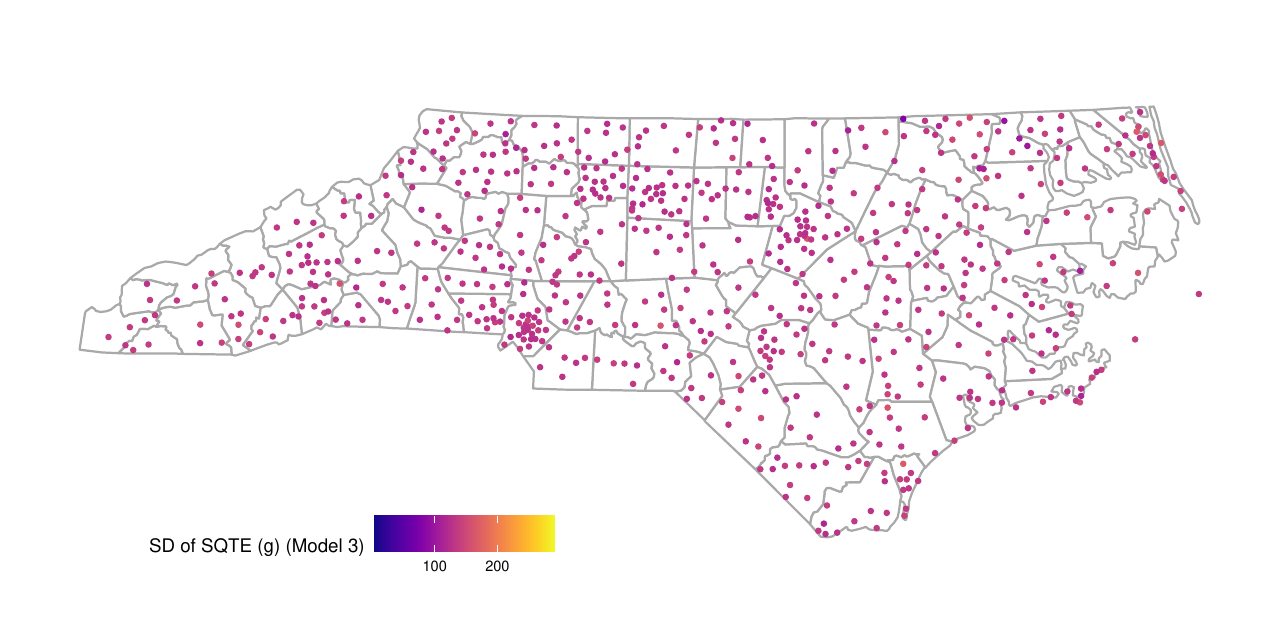}
    \caption{Estimated SQTE and standard deviation (SD) (in grams) of maternal smoking on the birth weight of newborn babies at $\tau = 0.05$ from Model 3 at all locations in North Carolina.}
\end{figure}
\begin{figure}[H]
\centering
\includegraphics[width=1\linewidth]{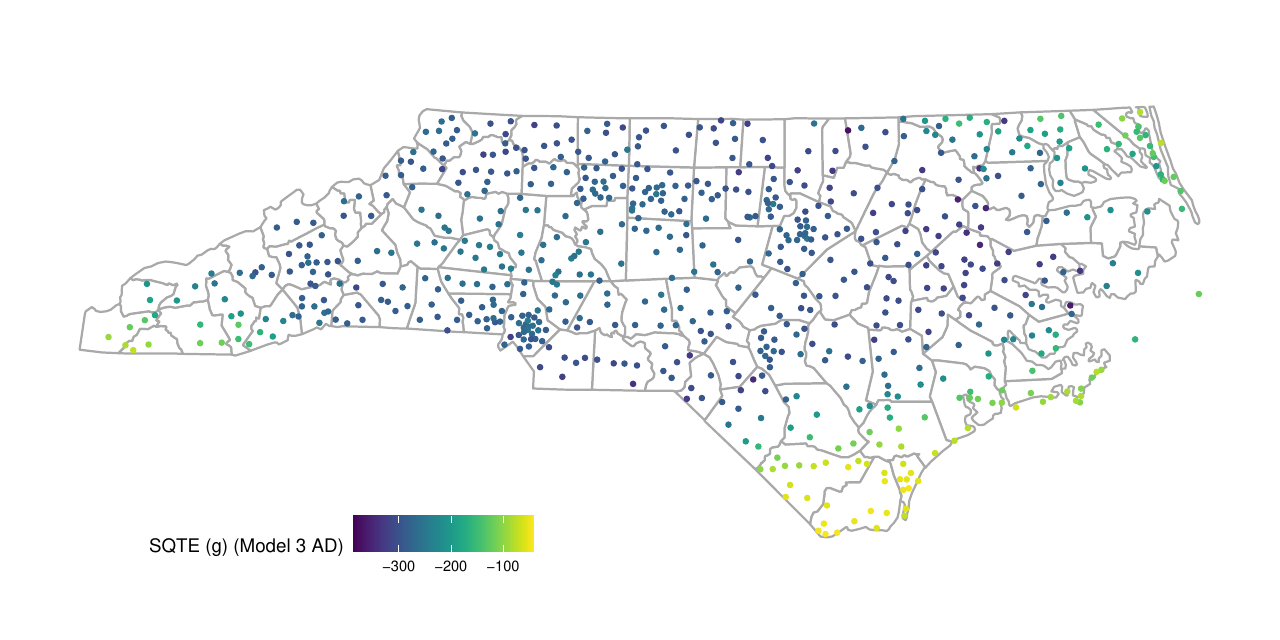}
  \includegraphics[width=1\linewidth]{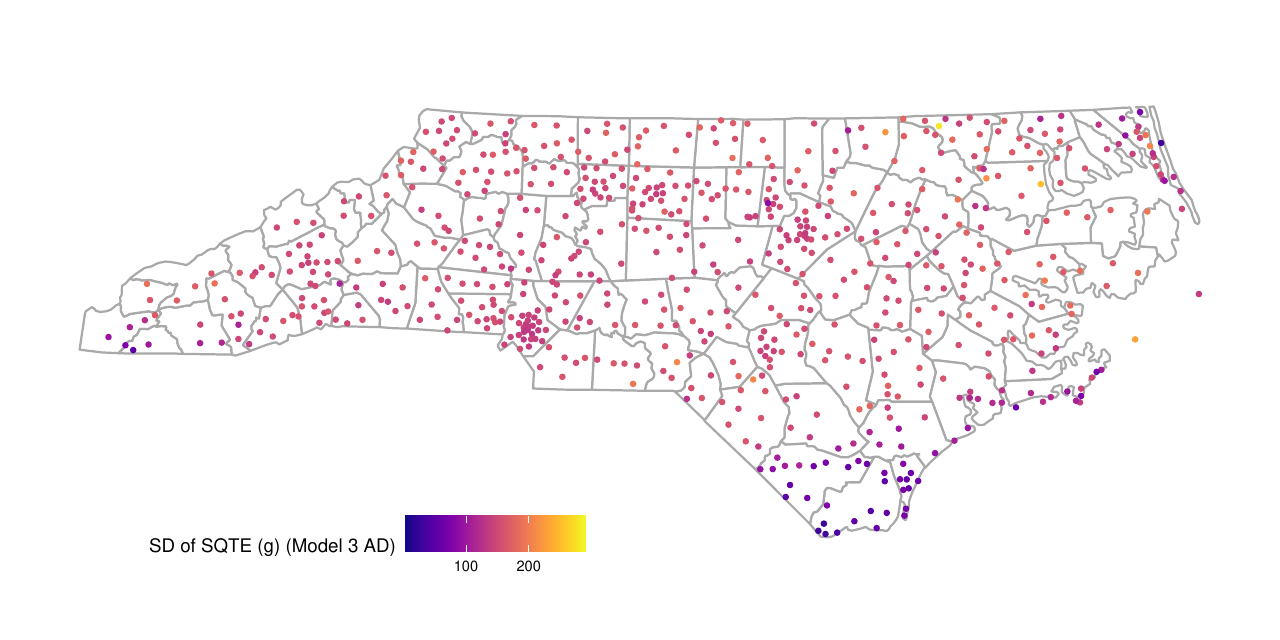}
    \caption{Estimated SQTE and standard deviation (SD) (in grams) of maternal smoking on the birth weight of newborn babies at $\tau = 0.05$ from Model 3 AD at all locations in North Carolina.}
\end{figure}

\begin{figure}[H]
	\centering
 \includegraphics[width=1\linewidth]{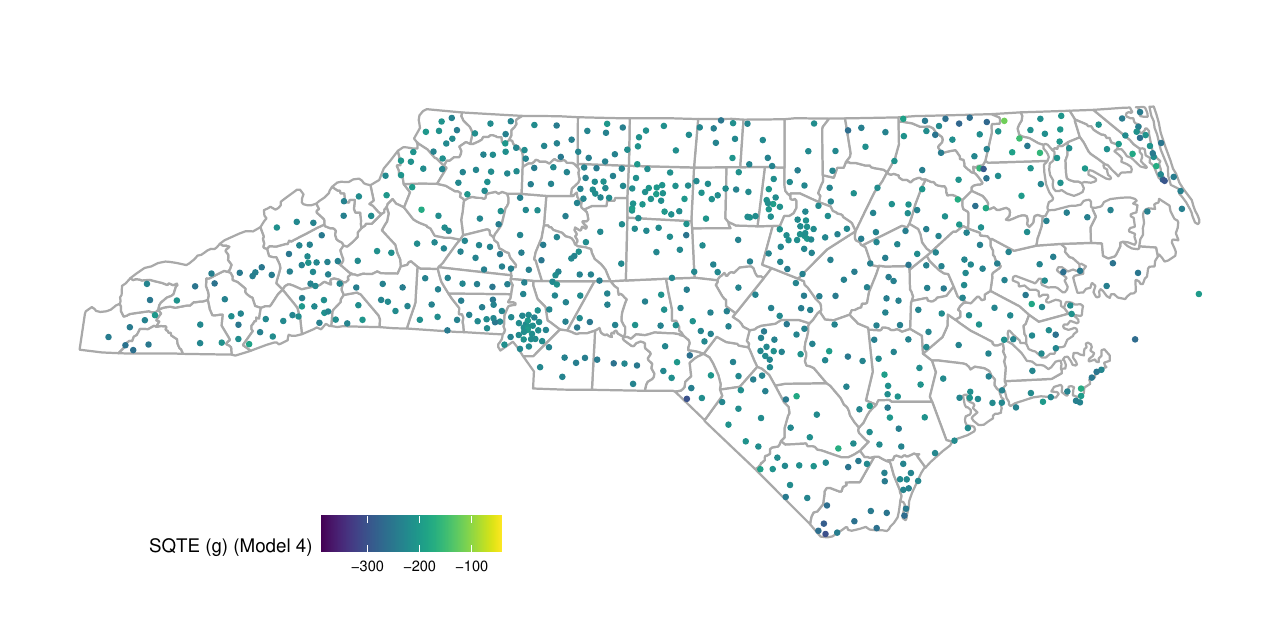}
  \includegraphics[width=1\linewidth]{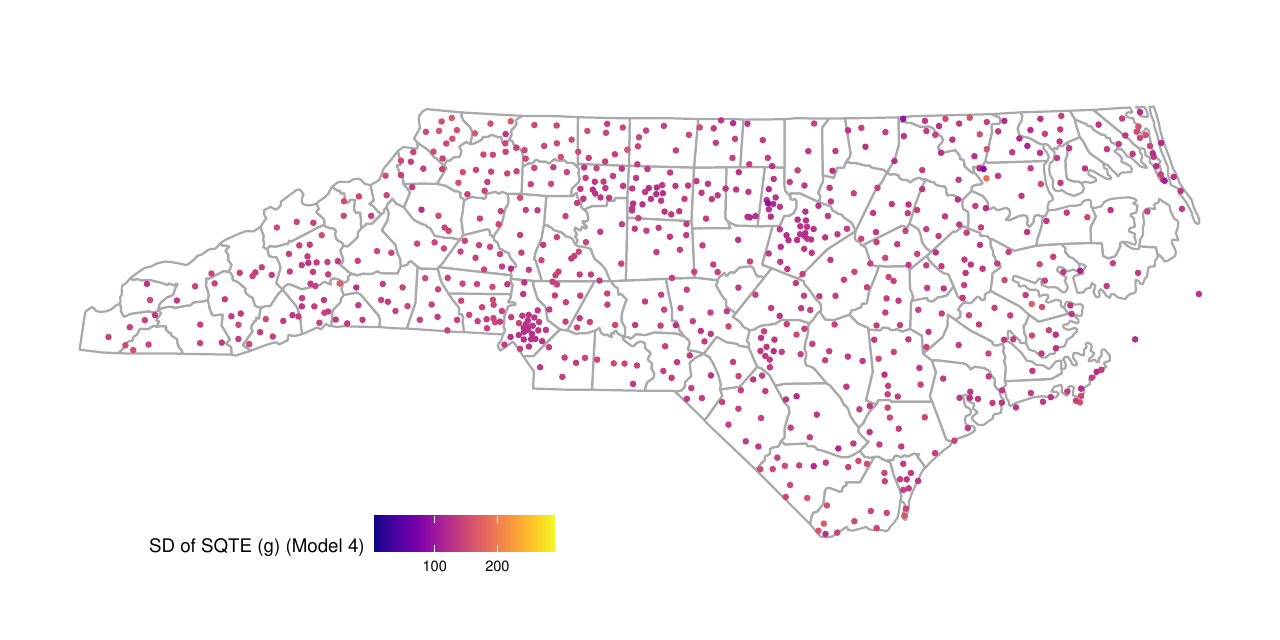}
    \caption{Estimated SQTE and standard deviation (SD) (in grams) of maternal smoking on the birth weight of newborn babies at $\tau = 0.05$ from Model 4 at all locations in North Carolina.}
\end{figure}
\begin{figure}[H]
\centering
\includegraphics[width=1\linewidth]{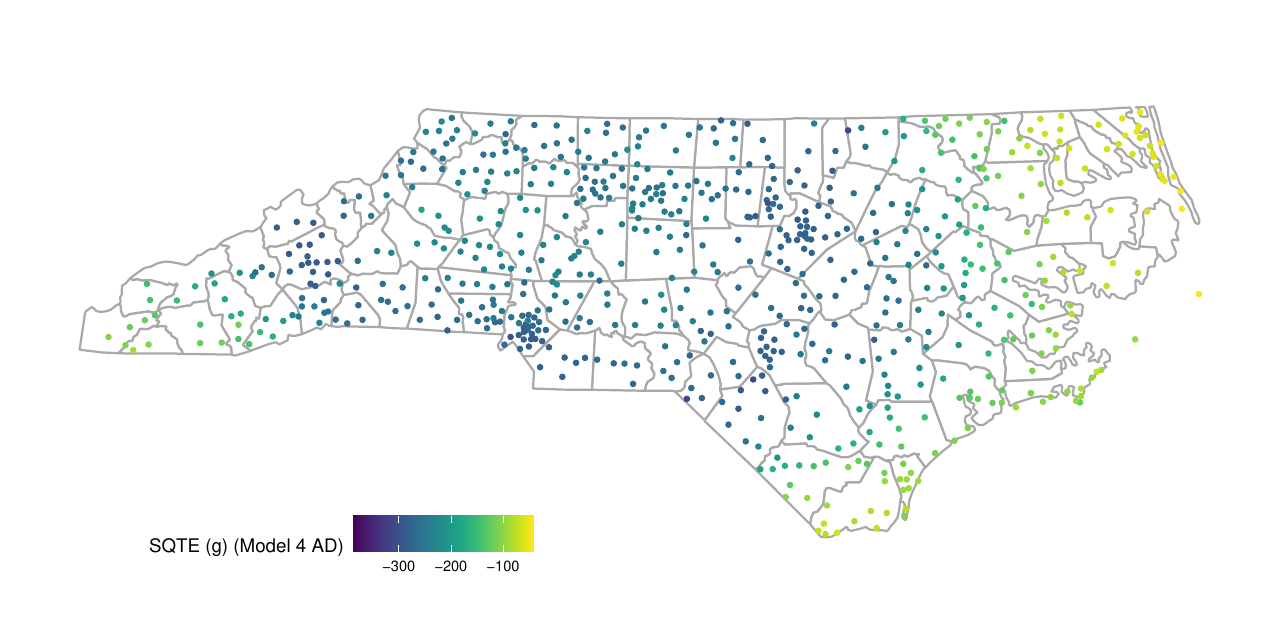}
  \includegraphics[width=1\linewidth]{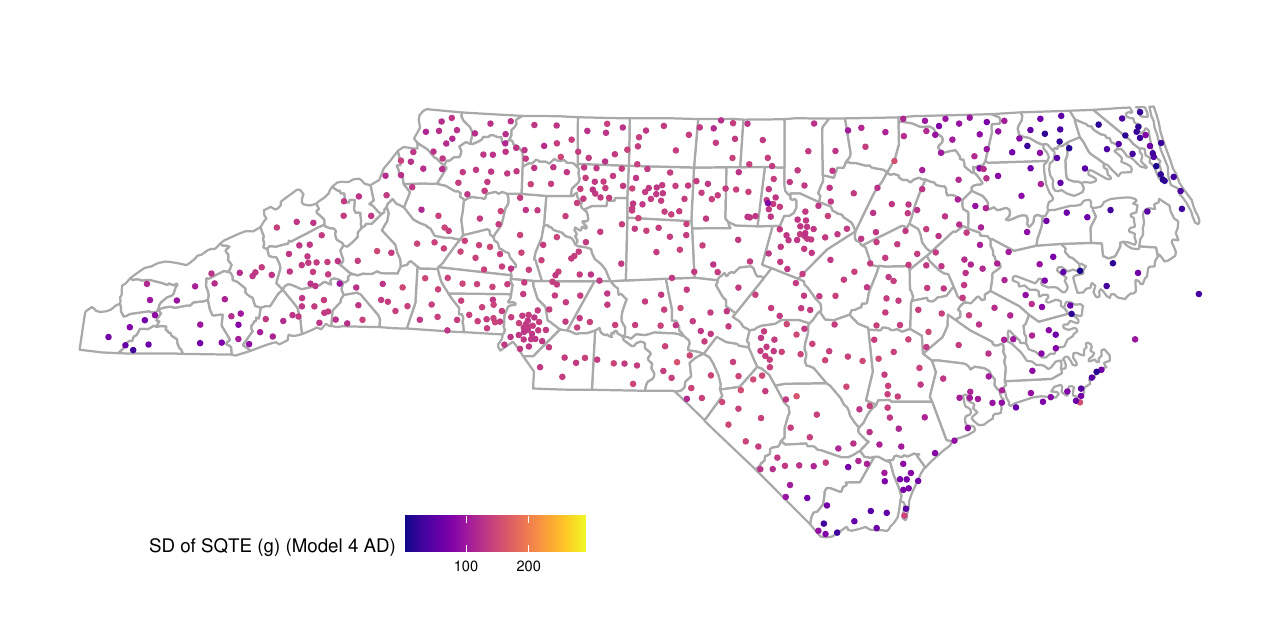}
    \caption{Estimated SQTE and standard deviation (SD) (in grams) of maternal smoking on the birth weight of newborn babies at $\tau = 0.05$ from Model 4 AD at all locations in North Carolina.}
\end{figure}

\begin{figure}[H]
	\centering
 \includegraphics[width=1\linewidth]{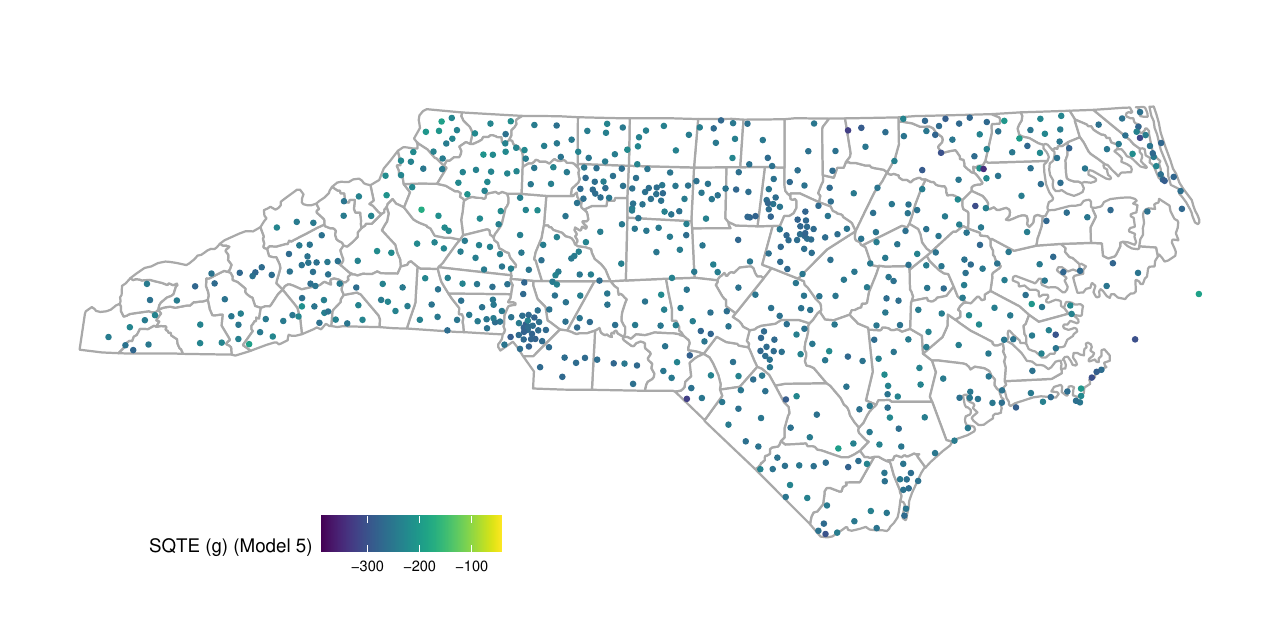}
  \includegraphics[width=1\linewidth]{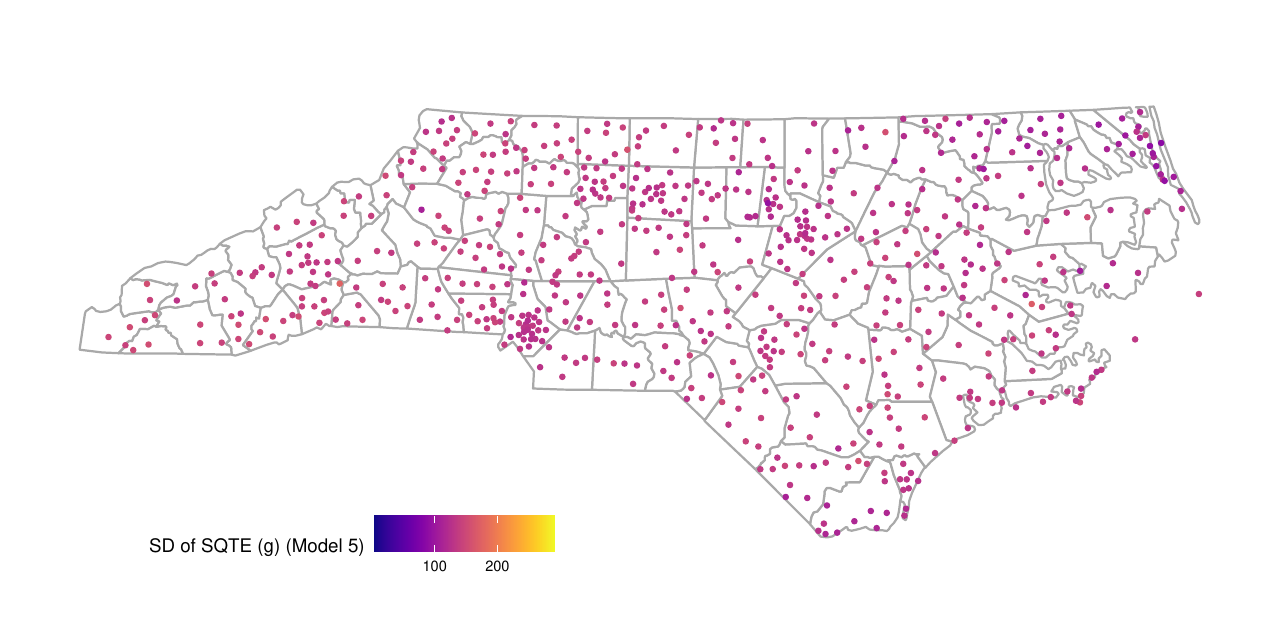}
    \caption{Estimated SQTE and standard deviation (SD) (in grams) of maternal smoking on the birth weight of newborn babies at $\tau = 0.05$ from Model 5 at all locations in North Carolina.}
\end{figure}
\begin{figure}[H]
\centering
\includegraphics[width=1\linewidth]{app001new_map_m5_mean_match.pdf}
  \includegraphics[width=1\linewidth]{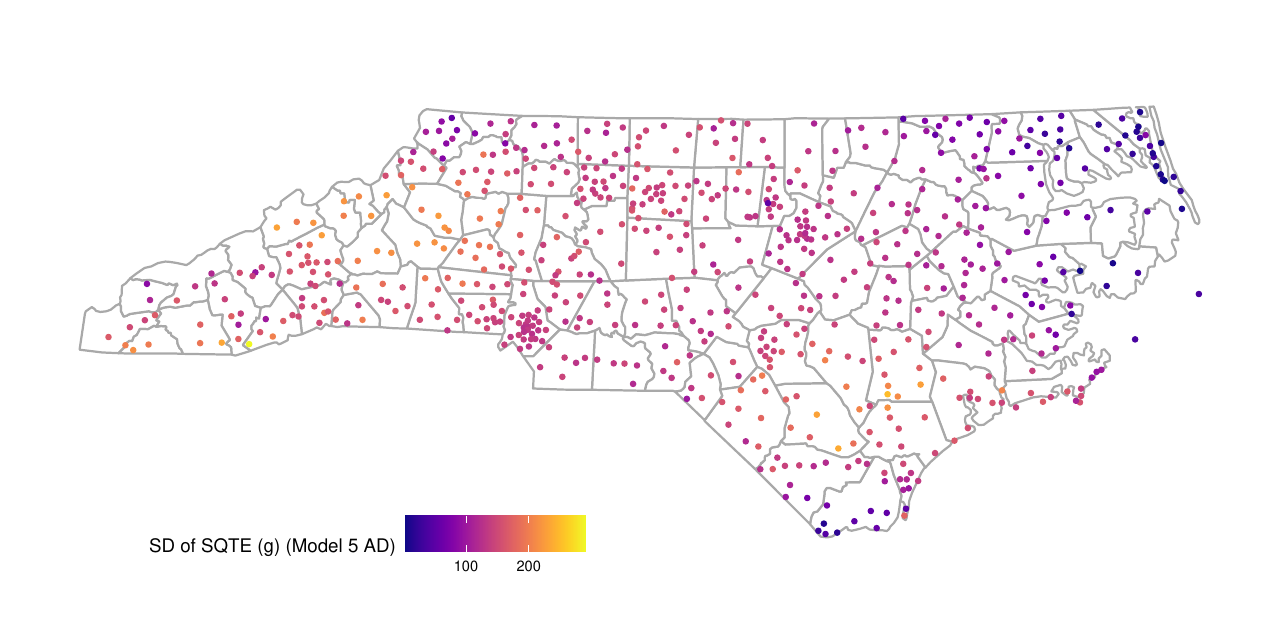}
    \caption{Estimated SQTE and standard deviation (SD) (in grams) of maternal smoking on the birth weight of newborn babies at $\tau = 0.05$ from Model 5 AD at all locations in North Carolina.}
\end{figure}

\end{document}